\documentclass{ws-ijmpa}

\usepackage{multirow}

\allowdisplaybreaks

\begin{document}

\markboth{Shu Luo $\&$ Zhi-zhong Xing} {Theoretical Overview on the Flavor
Issues of Massive Neutrinos}

\catchline{}{}{}{}{}
%
%%%%%%%%%%%%%%%%%%%%%%%%%%%%%%%%%%%%%%%%%%%%%%%%%%%%%%%%%%%%%%%%%%%%

\title{Theoretical Overview on the Flavor Issues of Massive Neutrinos}

\author{SHU LUO}
\address{Department of Physics and Institute of Theoretical
Physics and Astrophysics, \\
Xiamen University, Xiamen, Fujian, 361005 China \\
luoshu@xmu.edu.cn}

\author{ZHI-ZHONG XING}
\address{Institute of High Energy Physics
and Theoretical Physics Center for Science Facilities, \\ Chinese
Academy of Sciences, Beijing 100049, China; \\
Center for High Energy Physics, Peking University, Beijing 100080, China \\
xingzz@ihep.ac.cn}

\maketitle

%\begin{history}
%\received{\today}
%\revised{Day Month Year}
%\end{history}

\begin{abstract}
We present an overview on some basic properties of massive neutrinos
and focus on their flavor issues, including the mass spectrum,
flavor mixing pattern and CP violation. The lepton flavor structures
are explored by taking account of the observed value of the smallest
neutrino mixing angle $\theta^{}_{13}$. The impact of
$\theta^{}_{13}$ on the running behaviors of other flavor mixing
parameters is discussed in some detail. The seesaw-induced
enhancement of the electromagnetic dipole moments for three Majorana
neutrinos is also discussed in a TeV seesaw scenario.

\keywords{lepton flavor structure, neutrino mass, CP violation,
renormalization-group equation, electromagnetic dipole moment}
\end{abstract}

\ccode{PACS numbers: 11.25.Hf, 123.1K}

%\tableofcontents

\section{INTRODUCTION}

It is well known that two important experimental results were in the
news in 2012:
\begin{itemize}
\item On 8 March 2012, the Daya Bay Collaboration announced a $5.2 \sigma$
discovery of $\theta^{}_{13} \neq 0$ for this smallest neutrino mixing
angle \cite{DYB},
\begin{eqnarray}
\sin^2 2\theta^{}_{13} = 0.092 \pm 0.016 ({\rm stat}) \pm 0.005
({\rm syst}) ~~~~ (\pm 1\sigma ~ {\rm range}) \; ,
%     (1)
\end{eqnarray}
which is equivalent to $\theta^{}_{13} \simeq 8.8^\circ \pm
0.8^\circ$. The convincing Daya Bay result puts the preliminary
results of T2K \cite{T2K}, MINOS \cite{MINOS} and Double Chooz
\cite{Double Chooz}, which all hinted at $\theta^{}_{13} \neq 0$ in
2011, on solid ground. In particular, the fact that $\theta^{}_{13}$
is not strongly suppressed is a good news to the experimental
attempts towards a measurement of CP violation in the lepton sector.

\item On 4 July 2012, the ATLAS \cite{ATLAS}  and CMS \cite{CMS}
Collaborations at the Large Hadron Collider (LHC) independently announced
the discovery of a Higgs-like boson at the mass scale of 125 GeV to
127 GeV. If this result turns out to be true, it will have an important
impact on the development of neutrino physics because most of the neutrino
mass models depend on the existence of the Higgs particle(s) and
Yukawa interactions.
\end{itemize}
Therefore, a brief overview of where we are standing and where we are
expecting to go makes sense.

The remaining parts of this review paper are organized as follows.
In section 2 we give a fast overview of some fundamental neutrino
properties, such as the speed of neutrinos, the nature of massive
neutrinos and the number of neutrino species. Section 3 is devoted
to a brief description of the flavor issues of charged leptons and
neutrinos, including the mass spectrum, flavor mixing pattern and CP
violation. We compare the observed pattern of quark flavor mixing
with that of lepton flavor mixing. In section 4 we go into details
of possible lepton flavor structures by outlining two
phenomenological strategies and taking a number of typical examples.
The impact of large $\theta^{}_{13}$ on the running behaviors of
other flavor mixing parameters is discussed in section 5 by using
the one-loop renormalization-group equations (RGEs) in the framework
of the minimal supersymmetric standard model (MSSM). Section 6 is
devoted to the seesaw-enhanced electromagnetic dipole moments of
three Majorana neutrinos based on a TeV seesaw scenario. A summary
and some concluding remarks are given in section 7.

\section{IMMEDIATE QUESTIONS ON NEUTRINOS}

\subsection{Really Superluminal?}

The constancy of the speed of light $c$ in vacuum and the
independence of physical laws from the choice of inertial systems
are two fundamental propositions of the special relativity (SR)
\cite{Einstein}. If our world is Lorentz invariant, a free
particle's energy $E$, momentum ${\bm p}$ and rest mass $m$ satisfy
the relationship $\sqrt{E^{2}_{} - |{\bm p}|^{2}_{} c^{2}_{}} = m
c^{2}_{}$. The velocity of this particle turns out to be $v = c
\sqrt{1 - m^{2}_{} c^{4}_{} / E^{2}_{}}$, implying that it cannot
travel faster than light in vacuum. Could a particle be
superluminal? The answer would be yes if the particle had an
imaginary mass (called a ``tachyon'' \cite{tachyon}) or if the
Lorentz invariance were broken.

The OPERA Collaboration claimed a ``convincing" measurement of the
superluminal neutrinos in September 2011 \cite{OPERA}. But five months
later this story ended up with a mistake of the bad connection of the
optical fiber. The OPERA paper was updated in July 2012 by including
the new sources of errors, and the new result was in agreement
with the SR. Here let us quote Steven Weinberg's comments on the original
result of the OPERA experiment: ``The report of this experiment is
pretty impressive, but it bothers me that there is plenty of evidence
that all sorts of other particles never travel faster than light,
while observations of neutrinos are exceptionally difficult. It is as
if someone said that there are fairies in the bottom of their garden,
but they can only be seen on dark, foggy nights.''

An early measurement of the neutrino speed was done by using the pulsed pion
beams (produced by the pulsed proton beams hitting a target) at the
Fermilab in the 1970s \cite{Fermilab1,Fermilab2}. In this experiment
the speed of muons was compared with that of neutrinos and antineutrinos.
The same measurement was repeated in 2007 by using the MINOS detector
\cite{MINOS07}. In 2011 the speed of neutrinos was also measured in
a few other long-baseline neutrino experiments, such as the ICARUS
\cite{ICARUS1,ICARUS2}, Borexino \cite{Borexino} and LVD \cite{LVD}
experiments. But the most stringent constraint on the speed of neutrinos
was from the observational data of the Supernova 1987A
\cite{SN1987A1,SN1987A2,SN1987A3}: $|v - c| / c
\lesssim 10^{-9}$ obtained by comparing the arrival time of
light with that of neutrinos.
%%%%%%%%%%%%%%%%%%%%%%%% Fig. 1 %%%%%%%%%%%%%%%%%%%%%%%%
\begin{figure}
\centering
\includegraphics[width=1\textwidth]{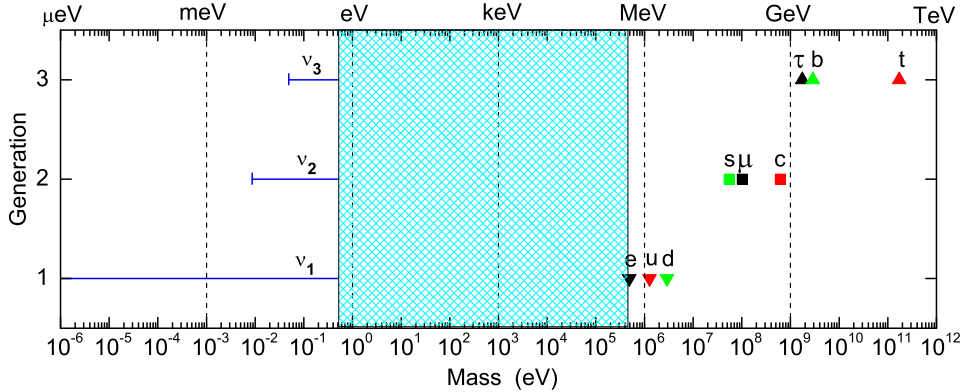}
\caption{A schematic illustration of the ``flavor hierarchy'' and
``flavor desert'' problems in the fermion mass spectrum at the
electroweak scale. Here the masses of three neutrinos are
assumed to have a normal hierarchy.}
\end{figure}
%%%%%%%%%%%%%%%%%%%%%%%%%%%%%%%%%%%%%%%%%%%%%%%%%%%%%%%%

\subsection{Definitely Massive?}

The neutrinos are massless in the standard model (SM) as a result
of its simple structure and renormalizability. On the one hand, the
SM does not contain any right-handed neutrinos, and thus there is no
way to write out the Dirac neutrino mass term. On the other hand,
the SM conserves the $SU(2)^{}_L$ gauge symmetry and only contains
the Higgs doublet, and thus the Majorana mass term is forbidden.
Although the SM accidently possesses the $(B-L)$ symmetry and
naturally allows neutrinos to be massless, the vanishing of
neutrino masses in the SM is not guaranteed by any fundamental
symmetry or conservation law. Today we have achieved a lot of robust
evidence for neutrino oscillations from solar, atmospheric, reactor
and accelerator neutrino experiments. The phenomenon of neutrino
oscillations implies that at least two of the three neutrinos must be
massive and the lepton flavors must be mixed. This is the first convincing
evidence for new physics beyond the SM.

Fig. 1 is a schematic plot of the mass spectrum of the SM
leptons and quarks at the electroweak scale. One can see that the
span between $m^{}_{1}$ and $m^{}_{t}$ is at least twelve orders of
magnitude. Furthermore, there exists an obvious ``desert''
spanning six orders of magnitude between the neutrino masses and
the masses of the charged fermions. Why do the SM fermions have
such hierarchy and desert puzzles? The
answer to this important question remains open. In particular,
the tiny neutrino masses must have a peculiar origin
(e.g., via the seesaw mechanisms \cite{SS1,SS2,SS3,SS4,SS5}).
Moreover, there might exist one or more keV sterile neutrinos in
the desert as a natural candidate for warm dark matter
\cite{keV1,keV2,keV3,keV4,keV5,keV6,keV7}.

\subsection{Dirac or Majorana?}

A {\it pure} Dirac mass term added into the SM is in general
disfavored, unless the theory is built by introducing extra
dimensions. Such a mass term in a renormalizable model of
electroweak interactions would worsen the fermion mass hierarchy
problem. An effective Majorana mass term given by the right-handed
neutrinos and their charge-conjugated counterparts is not forbidden
by the SM gauge symmetry, unless the contrived assumption of
lepton number conservation is imposed on the theory. Hence most
theorists believe that massive neutrinos are more likely to be
the Majorana particles and their salient feature is lepton number
violation.

The unique window to verify the Majorana nature of massive neutrinos
is to observe the neutrinoless doube-beta ($0\nu \beta\beta$) decay.
So far we have not obtained very convincing evidence for this
lepton-number-violating process. Even if the $0\nu \beta\beta$ decay
were never observed, one would still be unable to conclude that
massive neutrinos are the Dirac particles \cite{Xing03,Xing04}.
The effective mass of the $0\nu \beta\beta$ decay could vanish
if the Majorana CP-violating phases lie in some specific regions.
On the other hand, there are some other mechanisms
which can lead to the $0\nu \beta\beta$ decay. Such new physics
effects could be of the same order as or
even larger than the standard light-neutrino-exchange effect \cite{Rodejohann11,Rodejohann12}.

Given the SM interactions, a massive Dirac neutrino can have a
tiny (one-loop) magnetic dipole moment
$\mu^{}_\nu \sim 3\times 10^{-20} \mu^{}_{\rm B} (m^{}_\nu/0.1 ~
{\rm eV})$, where $\mu^{}_{\rm B}$ is the Bohr magneton
\cite{Shrock77,Shrock80}. In contrast, a massive Majorana neutrino
cannot have magnetic and electric dipole moments, because its
antiparticle is just itself. Both Dirac and Majorana neutrinos can
have {\it transition} dipole moments (of a size comparable with
$\mu^{}_\nu$) \cite{Vissani}, which may give rise to neutrino
decays, scattering effects with electrons, interactions with
external magnetic fields (red-giant stars, the sun, supernovae, and
so on), and contributions to neutrino masses. Current experimental
bounds on the neutrino dipole moments are at the level of $\mu^{}_\nu <
{\rm a ~ few} \times 10^{-11} \mu^{}_{\rm B}$.

\subsection{More than Three Species?}

It is well known that ``three" is a mystically popular number in
particle physics, such as three $Q=+2/3$ quarks, three $Q=-1/3$ quarks,
three $Q=-1$ leptons, three $Q=0$ neutrinos, three colors and three
forces in the SM. In this case, why do people want to go beyond
$N^{}_\nu = 3$?

The study of light sterile neutrinos has become a popular
direction in neutrino physics \cite{sterile review}. One is
motivated to consider such ``exotic" particles for several reasons.
On the theoretical side, the type-I seesaw mechanism
\cite{SS1,SS2,SS3,SS4,SS5} provides a very elegant interpretation of the
small masses of $\nu^{}_i$ (for $i=1,2,3$) with the help of two or
three heavy sterile neutrinos, and the latter can even help account
for the observed matter-antimatter asymmetry of the Universe via the
leptogenesis mechanism \cite{FY}. On the experimental side, the LSND
\cite{LSND}, MiniBooNE \cite{MiniBooNE} and reactor \cite{reactor}
antineutrino anomalies can all be explained as the active-sterile
antineutrino oscillations in the assumption of one or two species of
sterile antineutrinos whose masses are below 1 eV
\cite{Schwetz,Giunti}. Furthermore, a careful analysis of the
existing data on the Big Bang nucleosynthesis \cite{Mangano} or the
cosmic microwave background anisotropy, galaxy clustering and
supernovae Ia \cite{Raffelt1,Raffelt2,Raffelt3} seems to favor at
least one species of sterile neutrinos at the sub-eV mass scale. On
the other hand, sufficiently long-lived sterile neutrinos in the keV
mass range might serve for a good candidate for warm dark matter if
they were present in the early Universe \cite{Bode}.

If the three known neutrinos have mixing with a few new
degrees of freedom above or far above the Fermi scale, an exciting window
will be open to new physics at high energy scales. In this case,
however, the mixing between light and heavy neutrinos violates the
unitarity of the $3\times 3$ light neutrino mixing matrix and might
result in some observable effects in the future precision neutrino
experiments \cite{Xing3+3}.

\section{NEUTRINO MASSES AND FLAVOR MIXING}

There are three central concepts in flavor physics: mass, flavor
mixing and CP violation \cite{Xing04}. The phenomenon of lepton flavor
mixing at low energies is effectively described by a $3\times 3$
matrix $U$, the so-called Maki-Nakagawa-Sakata-Pontecorvo (MNSP)
matrix \cite{MNS1,MNS2}.
Given the unitarity of $U$, it can be parametrized in terms of three
angles and three phases \cite{PDG}:
\begin{eqnarray}
U = \left( \begin{matrix} c^{}_{12}
c^{}_{13} & s^{}_{12} c^{}_{13} & s^{}_{13} e^{-i\delta} \cr
-s^{}_{12} c^{}_{23} - c^{}_{12} s^{}_{13} s^{}_{23} e^{i\delta} &
c^{}_{12} c^{}_{23} - s^{}_{12} s^{}_{13} s^{}_{23} e^{i\delta} &
c^{}_{13} s^{}_{23} \cr s^{}_{12} s^{}_{23} - c^{}_{12} s^{}_{13}
c^{}_{23} e^{i\delta} & -c^{}_{12} s^{}_{23} - s^{}_{12} s^{}_{13}
c^{}_{23} e^{i\delta} & c^{}_{13} c^{}_{23} \cr \end{matrix} \right)
P^{}_\nu \; ,
%     (2)
\end{eqnarray}
where $c^{}_{ij} \equiv \cos\theta^{}_{ij}$, $s^{}_{ij} \equiv
\sin\theta^{}_{ij}$ (for $ij = 12, 13, 23$), and $P^{}_\nu ={\rm
Diag}\{e^{i\rho}, e^{i\sigma}, 1\}$ is physically relevant if
massive neutrinos are the Majorana particles.

Fogli {\it et al} \cite{Fogli} have recently done a global analysis
of current neutrino oscillation data and obtained the ranges of two
neutrino mass-squared differences ($\delta m^2 \equiv m^2_2 - m^2_1$
and $\Delta m^2 \equiv |m^2_3 - (m^2_1 + m^2_2)/2|$) and three
neutrino mixing angles, as listed in Table 1, where NH and IH
stand for the normal hierarchy ($m^{}_{1} < m^{}_{2} < m^{}_{3}$)
and the inverted hierarchy ($m^{}_{3} < m^{}_{1} < m^{}_{2}$),
respectively.
%%%%%%%%%%%%%%%%%%%%%%%%%%% Table 1 %%%%%%%%%%%%%%%%%%%%%%%%%%%%
\begin{table}
\tbl{Results of the global $3\nu$ oscillation analysis by Fogli
{\it et al} in 2012, including the best-fit values and allowed $1\sigma$,
$2\sigma$ and $3\sigma$ ranges for the neutrino oscillation parameters.} {\begin{tabular}{lcccc} \hline Parameter
& Best fit & $1\sigma$ range & $2\sigma$ range & $3\sigma$ range \\
\hline%---------------------------------------------------------------------
$\delta m^2/10^{-5}~\mathrm{eV}^2 $ (NH or IH) & 7.54
& 7.32 -- 7.80 & 7.15 -- 8.00 & 6.99 -- 8.18 \\
\hline%---------------------------------------------------------------------
$\sin^2 \theta_{12}/10^{-1}$ (NH or IH) & 3.07 & 2.91 -- 3.25
& 2.75 -- 3.42 & 2.59 -- 3.59 \\
\hline%---------------------------------------------------------------------
$\Delta m^2/10^{-3}~\mathrm{eV}^2 $ (NH) & 2.43 & 2.33 -- 2.49
& 2.27 -- 2.55 & 2.19 -- 2.62 \\
$\Delta m^2/10^{-3}~\mathrm{eV}^2 $ (IH) & 2.42 & 2.31 -- 2.49
& 2.26 -- 2.53 & 2.17 -- 2.61 \\
\hline%---------------------------------------------------------------------
$\sin^2 \theta_{13}/10^{-2}$ (NH) & 2.41 & 2.16 -- 2.66
& 1.93 -- 2.90 & 1.69 -- 3.13 \\
$\sin^2 \theta_{13}/10^{-2}$ (IH) & 2.44 & 2.19 -- 2.67
& 1.94 -- 2.91 & 1.71 -- 3.15 \\
\hline%---------------------------------------------------------------------
$\sin^2 \theta_{23}/10^{-1}$ (NH) & 3.86 & 3.65 -- 4.10
& 3.48 -- 4.48 &                       3.31 -- 6.37 \\
$\sin^2 \theta_{23}/10^{-1}$ (IH) & 3.92 & 3.70 -- 4.31
& 3.53 -- 4.84 $\oplus$ 5.43 -- 6.41 & 3.35 -- 6.63 \\
\hline%---------------------------------------------------------------------
$\delta/\pi$ (NH) & 1.08 & 0.77 -- 1.36 & --- & --- \\
$\delta/\pi$ (IH) & 1.09 & 0.83 -- 1.47 & --- & --- \\
\hline
\end{tabular}}
\end{table}
%%%%%%%%%%%%%%%%%%%%%%%%%%%%%%%%%%%%%%%%%%%%%%%%%%%%%%%%%%%%%%%%%

\subsection{Neutrino Mass Spectrum}

The two mass-squared differences of three neutrinos have been
determined, to a very good degree of accuracy, from current experimental
data: $\Delta m^2_{21} = \delta m^2 \approx 7.5 \times 10^{-5} ~ {\rm eV}^2$
and $\Delta m^2_{32} \approx \Delta m^2
\approx \pm 2.4 \times 10^{-3} ~ {\rm eV}^2$. The
absolute neutrino mass scale remains unknown and may hopefully be
determined in the following experimental or observational ways: the single
$\beta$ decay, the $0\nu \beta\beta$ decay and the cosmological
constraints. Fig. 2 shows the parameter space of
$\Sigma \equiv m^{}_1 + m^{}_2 + m^{}_3$,
$m^{}_{\beta} = \sqrt{m^{2}_1 |U^{}_{e1}|^2 + m^{2}_2 |U^{}_{e2}|^2 +
m^{2}_3 |U^{}_{e3}|^2}$ (the effective electron neutrino mass in the
$\beta$ decay) and
$m^{}_{\beta\beta} = |m^{}_1 U^2_{e1} + m^{}_2 U^2_{e2} + m^{}_3 U^2_{e3}|$
(the effective mass of the $0\nu \beta\beta$ decay).
A precision measurement of $m^{}_\beta$ and $\Sigma$ in the sub-eV range
could determine the neutrino mass hierarchy. In the
two lower panels of Fig. 2 there remains a large vertical spread in
the allowed slanted bands, as a result of the unknown Majorana
CP-violating phases in the $m^{}_{\beta\beta}$ components. This observation
indicates that more precise data in either the
$(m^{}_{\beta\beta},\,m^{}_\beta)$ plane or the
$(m^{}_{\beta\beta},\,\Sigma)$ plane might provide some useful constraints
on the Majorana phases.
%%%%%%%%%%%%%%%%%%%%%%%% Fig. 2 %%%%%%%%%%%%%%%%%%%%%%%%
\begin{figure}
\centering
\includegraphics[bb = 180 40 400 520,scale=0.6]{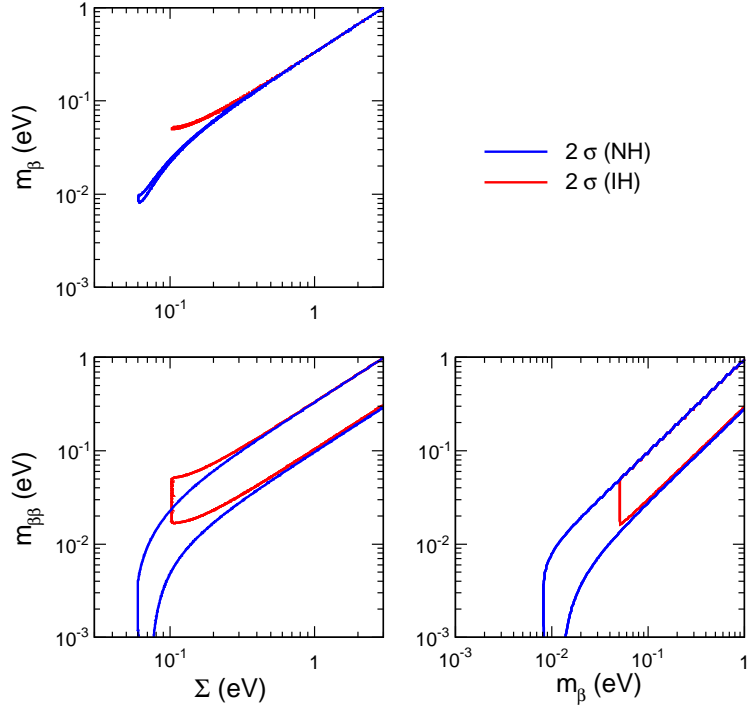}
\caption{Constraints obtained by Fogli {\it et al} in 2012
(at the $2\sigma$ level) in the planes charted by any two among the
absolute mass observables $m^{}_\beta$ (the effective mass of the
$\beta$ decay), $m^{}_{\beta\beta}$ (the effective mass of the
$0\nu\beta\beta$ decay) and $\Sigma$ (the sum of three
neutrino masses). The blue (red) bands refer to the normal (inverted)
neutrino mass hierarchy.}
\end{figure}
%%%%%%%%%%%%%%%%%%%%%%%%%%%%%%%%%%%%%%%%%%%%%%%%

Before the absolute mass scale is determined, there remain two
open questions: (1) is $m^{}_3$ bigger or smaller than $m^{}_1$?
(2) can one neutrino mass ($m^{}_1$ or $m^{}_3$) be vanishing or
vanishingly small? The first question
awaits an experimental answer in the foreseeable future, such as a
long-baseline neutrino oscillation experiment with appreciable
terrestrial matter effects \cite{GG} or a long-baseline
reactor antineutrino oscillation experiment with accurate
information on the energy spectrum \cite{hierarchy1,hierarchy2}.
A theoretical answer to the second question is strongly
model-dependent. Examples of this type include the minimal type-I seesaw
mechanism with two heavy Majorana neutrinos \cite{MSM1,MSM2} or the
Friedberg-Lee ansatz with an effective Dirac or Majorana neutrino
mass operator \cite{FL1,FL2,FL3,FL4,FL5,FL6,FL7}.

\subsection{Flavor Mixing Pattern}

The fact that the smallest neutrino mixing angle $\theta^{}_{13}$ is not
strongly suppressed leads us to some new questions about the
feature of lepton flavor mixing: (1) Can the  relatively large
$\theta^{}_{13}$ be understood by an underlying flavor symmetry or is it
generated by a symmetry breaking mechanism or quantum corrections?
(2) Does $\theta^{}_{23} \simeq 45^\circ$ still hold? (3) What is the
strength of leptonic CP violation?

The structure of the MNSP lepton flavor mixing matrix $U$
is significantly different from that of the
Cabibbo-Kobayashi-Maskawa (CKM) quark flavor
mixing matrix $V$. The CKM matrix is nearly the unit
matrix up to some small corrections, while the MNSP
matrix has an approximate $\mu$-$\tau$ symmetry. The full
$\mu$-$\tau$ symmetry of $U$ in modulus is described by the
equalities
\begin{eqnarray}
|U^{}_{\mu 1}| = |U^{}_{\tau 1}| \; , ~~~~
|U^{}_{\mu 2}| = |U^{}_{\tau 2}| \; , ~~~~
|U^{}_{\mu 3}| = |U^{}_{\tau 3}| \; ,
%     (3)
\end{eqnarray}
equivalent to two independent sets of conditions in the standard
parametrization given in Eq. (2) \cite{XZ08}:
\begin{eqnarray}
\theta^{}_{23} = 45^\circ \; , ~~~ \theta^{}_{13} = 0^\circ \; ,
%     (4)
\end{eqnarray}
or
\begin{eqnarray}
\theta^{}_{23} = 45^\circ \; ,
~~~ \delta = \pm 90^\circ \; .
%     (5)
\end{eqnarray}
If $\theta^{}_{23}$ is exactly equal to $45^\circ$, then one may
arrive at a partial $\mu$-$\tau$ permutation symmetry in the MNSP
matrix $U$ (i.e., the equality $|U^{}_{\mu 3}| = |U^{}_{\tau 3}|$).

Now that $\theta^{}_{13} \neq 0^\circ$ has firmly been established
by the Daya Bay experiment \cite{DYB}, it becomes crucial to check
the deviation of $\theta^{}_{23}$ from $45^\circ$ and (or) a
possible departure of $\delta$ from $\pm 90^\circ$. We speculate
that $U$ might have an approximate $\mu$-$\tau$ symmetry with
$|U^{}_{\mu i}| \simeq |U^{}_{\tau i}|$, in contrast with the
approximate off-diagonal symmetry of the CKM matrix $V$ in modulus
(i.e., $|V^{}_{us}| \simeq |V^{}_{cd}|$, $|V^{}_{cb}| \simeq
|V^{}_{ts}|$ and $|V^{}_{ub}| \simeq |V^{}_{td}|$ \cite{PDG}).

In the basis where the flavor eigenstates of three charged leptons
are identified with their mass eigenstates (i.e., $M^{}_l =
\widehat{M}^{}_l$), the Majorana neutrino mass matrix of the form
\begin{eqnarray}
M^{}_\nu = \left( \begin{matrix} a & b & -b \cr b &
\hspace{0.17cm} c \hspace{0.17cm} & d \cr -b & d & c \cr \end{matrix}
\right) \;
%     (6)
\end{eqnarray}
predicts the $\mu$-$\tau$ permutation symmetry of the MNSP matrix
$U$ with $\theta^{}_{13} = 0^\circ$ and $\theta^{}_{23} = 45^\circ$;
while the mass matrix of the form
\begin{eqnarray}
M^{}_\nu = \left( \begin{matrix} a & b & - b^* \cr b & c & d \cr
- b^* & d & c^* \cr \end{matrix} \right) \;
%     (7)
\end{eqnarray}
leads us to the $\mu$-$\tau$ symmetry of $U$ with $\delta = \pm
90^\circ$ and $\theta^{}_{23} = 45^\circ$. In either of the above
textures of $M^{}_\nu$, its entries have certain kinds of linear
correlations or equalities and thus can be generated from some
underlying flavor symmetries. In view of the experimental evidence
for $\theta^{}_{13} \neq 0^\circ$ \cite{DYB}, the pattern of
$M^{}_\nu$ in Eq. (6) has to be modified. For a similar reason, the
more reliable and accurate experimental knowledge on
$\theta^{}_{23}$ and $\delta$ will be useful for us to
identify the effect of $\mu$-$\tau$ symmetry breaking and build more
realistic models for lepton mass generation, flavor mixing and CP
violation.

\subsection{CP and T Violation}

If neutrinos are the Majorana particles, the $3\times 3$
MNSP matrix $U$ contains three CP-violating phases
$\delta$, $\rho$ and $\sigma$. Among them, $\delta$ determines the
strength of CP and T violation in neutrino oscillations, because
both $P(\nu^{}_\alpha \to \nu^{}_\beta) - P(\overline{\nu}^{}_\alpha
\to \overline{\nu}^{}_\beta)$ and $P(\nu^{}_\alpha \to \nu^{}_\beta)
- P(\nu^{}_\beta \to \nu^{}_\alpha)$ are proportional to the leptonic
Jarlskog invariant ${\cal J}^{}_{l} = \sin\theta^{}_{12}
\cos\theta^{}_{12} \sin\theta^{}_{23} \cos\theta^{}_{23}
\sin\theta^{}_{13} \cos^2\theta^{}_{13} \sin\delta$ in vacuum
\cite{J}. The phases $\rho$ and $\sigma$, which have
nothing to do with neutrino oscillations, are associated with the
$0\nu \beta\beta$ decay. Note that $\delta$ itself is also of the
Majorana nature, although it is usually referred to as the Dirac
phase: one reason is that $\delta$ may appear in other
lepton-number-violating processes, even if it can sometimes be arranged
{\it not} to appear in the $0\nu \beta\beta$ decay; and the other
reason is that $\delta$, $\rho$ and $\sigma$ are actually entangled
with one another in the RGE running from one energy scale to another.

The fact that $\theta^{}_{13}$ is not strongly suppressed is
certainly a good news to the experimental attempts towards a final
measurement of CP violation in the lepton sector. The reason is
simply that the strength of leptonic CP violation (i.e.,
${\cal J}^{}_{l}$) is proportional to $\sin\theta^{}_{13}$.
In the quark sector one has determined the
corresponding Jarlskog invariant ${\cal J}^{}_q \simeq 3\times
10^{-5}$ \cite{PDG} and attributed its smallness to the strongly
suppressed values of quark flavor mixing angles (i.e.,
$\vartheta^{}_{\rm C} \equiv \vartheta^{}_{12} \simeq 13^\circ$,
$\vartheta^{}_{13} \simeq 0.2^\circ$ and $\vartheta^{}_{23} \simeq
2.4^\circ$). In the lepton sector both $\theta^{}_{12}$ and
$\theta^{}_{23}$ are large, and thus it is possible to achieve a
relatively large value of ${\cal J}^{}_l$ if the CP-violating phase
$\delta$ is not small either. Taking $\theta^{}_{12} \simeq
34^\circ$, $\theta^{}_{13} \sim 9^\circ$ and $\theta^{}_{23} \simeq
45^\circ$ as a realistic example of $U$, we arrive at ${\cal J}^{}_l
\simeq 0.036 \sin\delta$, implying that the magnitude of leptonic CP
violation can actually reach the percent level in neutrino
oscillations if $\delta$ is not strongly
suppressed. Whether CP violation is significant or not turns out to
be an important question in lepton physics, especially in neutrino
phenomenology.

\subsection{Comparison between the MNSP and CKM Matrices}

The relative sizes of the nine elements of the MNSP matrix $U$
cannot be completely fixed unless we have known $\theta^{}_{23} >
45^\circ$ or $\theta^{}_{23} < 45^\circ$ as well as the range of
$\delta$. With the help of the available experimental data and the
unitarity of $U$, we find
\begin{eqnarray}
|U^{}_{e 1}| > |U^{}_{\mu 3}| \sim |U^{}_{\tau 3}| >
|U^{}_{\mu 2}| \sim |U^{}_{\tau 2}| > |U^{}_{e 2}| > |U^{}_{\mu 1}|
\sim |U^{}_{\tau 1}| > |U^{}_{e 3}| \; ,
%     (8)
\end{eqnarray}
where ``$\sim$" implies that the relative magnitudes of $|U^{}_{\mu
i}|$ and $|U^{}_{\tau i}|$ (for $i=1,2,3$) remain undetermined at
present. In comparison, the nine elements of the CKM matrix $V$ are
known to have the following hierarchy \cite{Xing96,Xing97}:
\begin{eqnarray}
|V^{}_{tb}| > |V^{}_{ud}| > |V^{}_{cs}| \gg
|V^{}_{us}| > |V^{}_{cd}| \gg |V^{}_{cb}| > |V^{}_{ts}|
\gg |V^{}_{td}| > |V^{}_{ub}| \; .
%     (9)
\end{eqnarray}
There is a striking similarity between the quark and
lepton flavor mixing matrices: the smallest elements of both $V$ and
$U$ appear in their respective top-right corners.

In the history of flavor physics it took quite a long time to
measure the four independent parameters of $V$, but
the experimental development had a clear roadmap:
\begin{eqnarray}
\vartheta^{}_{12} ~ ({\rm or} ~ |V^{}_{us}|) ~~ \Longrightarrow ~~
\vartheta^{}_{23} ~ ({\rm or} ~ |V^{}_{cb}|) ~~ \Longrightarrow ~~
\vartheta^{}_{13} ~ ({\rm or} ~ |V^{}_{ub}|) ~~ \Longrightarrow ~~
\delta ~({\rm quark}) \; .
%     (10)
\end{eqnarray}
Namely, the observation of the largest mixing angle
$\vartheta^{}_{12}$ was the first step, the determination of the
smallest mixing angle $\vartheta^{}_{13}$ was an important turning
point, and then the quark flavor physics entered an era of precision
measurements in which CP violation could be explored and new physics
could be searched for. Interestingly and hopefully, the lepton
flavor physics is repeating the same story:
\begin{eqnarray}
\theta^{}_{23} ~ ({\rm or} ~ |U^{}_{\mu 3}|) ~~ \Longrightarrow ~~
\theta^{}_{12} ~ ({\rm or} ~ |U^{}_{e 2}|) ~~ \Longrightarrow ~~
\theta^{}_{13} ~ ({\rm or} ~ |U^{}_{e 3}|) ~~ \Longrightarrow ~~
\delta ~({\rm lepton}) \; ,
%     (11)
\end{eqnarray}
where $\theta^{}_{23}$ is the largest and $\theta^{}_{13}$ is the
smallest. The observation of $\theta^{}_{13}$ in the Daya Bay experiment is
paving the way for future experiments to study leptonic CP violation
and to look for possible new physics (e.g., whether the $3\times 3$
MNSP matrix $U$ is exactly unitary or not \cite{NewX}),
in particular through
the measurements of neutrino oscillations for different sources of
neutrino beams. The Majorana nature of three massive neutrinos and
their other two CP-violating phases (i.e., $\rho$ and $\sigma$) can
also be probed in the new era of neutrino physics.

\section{POSSIBLE LEPTON FLAVOR STRUCTURES}

\subsection{Two Phenomenological Strategies}

The MNSP matrix $U$ actually describes a fundamental mismatch
between the flavor and mass eigenstates of six
leptons, or a mismatch between diagonalizations of
the charged-lepton mass matrix $M^{}_l$ and the effective neutrino
mass matrix $M^{}_\nu$ in a given model, no matter whether the
origin of neutrino masses is attributed to the seesaw mechanisms or
not \cite{FX00}. Assuming massive neutrinos to be the Majorana
particles, we may simply write out the leptonic mass terms as
\begin{eqnarray}
-{\cal L}^{}_{\rm mass} = \overline{ \left(\begin{matrix} e^\prime &
\mu^\prime & \tau^\prime \end{matrix} \right)^{}_{\rm L}} \ M^{}_l
\left(\begin{matrix} e^\prime \cr \mu^\prime \cr \tau^\prime
\end{matrix} \right)^{}_{\rm R} + \ \frac{1}{2} \ \overline{
\left(\begin{matrix} \nu^{}_e & \nu^{}_\mu & \nu^{}_\tau
\end{matrix} \right)^{}_{\rm L}} \ M^{}_\nu \left(\begin{matrix}
\nu^{c}_e \cr \nu^{c}_\mu \cr \nu^{c}_\tau \cr \end{matrix}
\right)^{}_{\rm R} + {\rm h.c.} \; ,
%     (12)
\end{eqnarray}
where ``$\prime$" stands for the flavor eigenstates of charged
leptons, ``$c$" denotes the charge-conjugated neutrino fields, and
$M^{}_\nu$ is symmetric. By using the unitary matrices $O^{}_l$,
$O^\prime_l$ and $O^{}_\nu$, one can diagonalize $M^{}_l$ and
$M^{}_\nu$ through the transformations $O^\dagger_l M^{}_l
O^\prime_l = \widehat{M}^{}_l \equiv {\rm Diag}\{m^{}_e, m^{}_\mu,
m^{}_\tau\}$ and $O^\dagger_\nu M^{}_\nu O^*_\nu =
\widehat{M}^{}_\nu \equiv {\rm Diag}\{m^{}_1, m^{}_2, m^{}_3 \}$,
respectively. Then one arrives at the lepton mass terms in terms of
the mass eigenstates:
\begin{eqnarray}
-{\cal L}^\prime_{\rm mass} = \overline{ \left(\begin{matrix} e &
\mu & \tau \end{matrix} \right)^{}_{\rm L}} \ \widehat{M}^{}_l
\left(\begin{matrix} e \cr \mu \cr \tau \end{matrix} \right)^{}_{\rm
R} + \ \frac{1}{2} \ \overline{ \left(\begin{matrix} \nu^{}_1 &
\nu^{}_2 & \nu^{}_3 \end{matrix} \right)^{}_{\rm L}} \
\widehat{M}^{}_\nu \left(\begin{matrix} \nu^{c}_1 \cr \nu^{c}_2 \cr
\nu^{c}_3 \cr \end{matrix} \right)^{}_{\rm R} + {\rm h.c.} \; .
%     (13)
\end{eqnarray}
Extending this basis transformation to the standard
charged-current interactions, we immediately obtain
\begin{eqnarray}
-{\cal L}^{}_{\rm cc} = \frac{g}{\sqrt{2}} \ \overline{
\left(\begin{matrix} e & \mu & \tau \end{matrix} \right)^{}_{\rm L}}
\ \gamma^\mu U \left(\begin{matrix} \nu^{}_1 \cr \nu^{}_2 \cr
\nu^{}_3 \cr \end{matrix} \right)^{}_{\rm L} W^-_\mu + {\rm h.c.} \; ,
%     (14)
\end{eqnarray}
in which $U = O^\dagger_l O^{}_\nu$. The
above treatment is most general at a given energy scale (e.g., the
electroweak scale). There are two different strategies of
phenomenologically understanding the structure of the MNSP matrix
\cite{Xing2012}.

(1) {\it The mixing angles of $U$ are associated with the lepton
mass ratios}.
The structure of lepton flavor mixing is directly determined by the
structures of $O^{}_l$ and $O^{}_\nu$. Since these two unitary
matrices are used to diagonalize $M^{}_l$ and $M^{}_\nu$,
respectively, their structures are governed by those of $M^{}_l$ and
$M^{}_\nu$, whose eigenvalues are the physical lepton masses.
Therefore, we anticipate that the dimensionless flavor mixing angles
of $U$ should be certain kinds of functions whose variables include
four independent mass ratios of three charged leptons and three
neutrinos. Namely,
\begin{eqnarray}
\theta^{}_{ij} = f\left(\frac{m^{}_\alpha}{m^{}_\beta},
\frac{m^{}_k}{m^{}_l}, \cdots\right) \; ,
%     (15)
\end{eqnarray}
where the Greek subscripts denote the charged leptons, the Latin
subscripts stand for the neutrinos, and ``$\cdots$" implies other
dimensionless parameters originating from the lepton mass
matrices. Such an expectation has proved valid in the quark sector
to explain why the relation $\sin\vartheta^{}_{\rm C} \simeq
\sqrt{m^{}_d/m^{}_s}$ works quite well and how the hierarchical
structure of the CKM matrix $V$ is related to the strong hierarchies
of quark masses (i.e., $m^{}_u \ll m^{}_c \ll m^{}_t$ and $m^{}_d
\ll m^{}_s \ll m^{}_b$) \cite{FX2000}. As for the phenomenon of
lepton flavor mixing, it is apparently difficult to link two large
mixing angles $\theta^{}_{12}$ and $\theta^{}_{23}$ to
$m^{}_e/m^{}_\mu \simeq 4.7 \times 10^{-3}$ and
$m^{}_\mu/m^{}_\tau \simeq 5.9 \times 10^{-2}$ \cite{XZZ1,XZZ2}.
Hence one may consider to ascribe the largeness of $\theta^{}_{12}$
and $\theta^{}_{23}$ to a weak hierarchy of three neutrino masses,
such as the conjecture $\tan\theta^{}_{12} \simeq
\sqrt{m^{}_1/m^{}_2}$ \cite{FX06,FX09}.

To establish a direct relation between $\theta^{}_{ij}$ and lepton
mass ratios, one has to specify the textures of $M^{}_l$ and
$M^{}_\nu$ by allowing some of their elements to vanish or to be
vanishingly small. A typical example of this kind is the
Fritzsch ansatz \cite{Fritzsch78,Fritzsch79},
\begin{eqnarray}
M^{}_{l,\nu} = \left( \begin{matrix} 0 & \times & 0 \cr
\times & 0 & \times \cr 0 & \times & \times \cr \end{matrix} \right) \; ,
%     (16)
\end{eqnarray}
which is able to account for the present neutrino oscillation data to an
acceptable degree of accuracy (e.g., $\sin\theta^{}_{23} \simeq
\sqrt{m^{}_\mu/m^{}_\tau} + \sqrt{m^{}_2/m^{}_3} \simeq 0.65$)
\cite{Xing02,ZX05,FXZ11}. Another well-known and
viable example is the two-zero textures of $M^{}_\nu$ in the basis
where $M^{}_l$ is diagonal \cite{Zero1,Zero2,Zero3,Zero4}. Note that
the texture zeros of a fermion mass matrix dynamically mean that the
corresponding matrix elements are sufficiently suppressed as
compared with their neighboring counterparts, and they can be
derived from a certain flavor symmetry in a given theoretical
framework (e.g., with the help of the Froggatt-Nielson mechanism
\cite{FN} or discrete flavor symmetries \cite{Grimus}).

(2) {\it The lepton flavor mixing matrix $U$ consists of a constant
leading term $U^{}_0$ and a small perturbation term $\Delta U$}.
In fact, $U$ has been conjectured to have the following structure
for a quite long time \cite{FX00,FX96,FX98}:
\begin{eqnarray}
U = \left(U^{}_0 + \Delta U\right) P^{}_\nu \; ,
%     (17)
\end{eqnarray}
in which the leading term $U^{}_0$ is a constant matrix responsible
for two larger mixing angles $\theta^{}_{12}$ and $\theta^{}_{23}$,
and the next-to-leading term $\Delta U$ is a perturbation
responsible for both the smallest mixing angle $\theta^{}_{13}$ and
the Dirac CP-violating phase $\delta$. So far a lot of flavor
symmetries have been brought into exercise to derive $U^{}_0$, while
$\Delta U$ might originate from either an explicit flavor symmetry
breaking scenario or some finite quantum corrections at a given
energy scale or from a superhigh-energy scale to the electroweak
scale.

In this case the MNSP matrix $U$ is approximately a constant matrix
whose mixing angles are independent of the lepton mass ratios. This
conjecture is actually in conflict with the conjecture made in the
first strategy.  The reason for this ``conflict" is rather simple:
the assumed structures of lepton flavor mixing in Eqs. (15) and (17)
correspond to two different structures of lepton mass matrices. As
we have pointed out above, the direct dependence of $\theta^{}_{ij}$
on $m^{}_\alpha/m^{}_\beta$ and $m^{}_k/m^{}_l$ is usually a direct
consequence of the texture zeros of $M^{}_l$ and (or) $M^{}_\nu$. In
contrast, a constant flavor mixing pattern $U^{}_0$ may arise from
some special textures of $M^{}_l$ and (or) $M^{}_\nu$ whose entries
have certain kinds of linear correlations or equalities. For
instance, the texture \cite{FL1,FL2,FL3,FL4,FL5,FL6,FL7}
\begin{eqnarray}
M^{}_\nu = \left( \begin{matrix} b+c & -b & -c \cr -b & a+b & -a \cr
-c & -a & a+c \cr \end{matrix} \right) \;
%     (18)
\end{eqnarray}
assures $O^{}_\nu$ to be of the tri-bimaximal mixing pattern to be
discussed in section 4.2. This
neutrino mass matrix has no zero entries, but its nine elements
satisfy the sum rules $(M^{}_\nu)^{}_{i1} + (M^{}_\nu)^{}_{i2} +
(M^{}_\nu)^{}_{i3} = 0$ and $(M^{}_\nu)^{}_{1j} + (M^{}_\nu)^{}_{2j}
+ (M^{}_\nu)^{}_{3j} = 0$ (for $i, j=1,2,3$). Such correlative
relations are similar to those texture zeros in the sense that both
of them may reduce the number of free parameters associated with
lepton mass matrices, making some predictions for the lepton flavor
mixing angles technically possible.

In short, one may try to understand the structure of the MNSP matrix $U$
by following two phenomenological strategies:
\begin{enumerate}
\item to explore
possible relations between the flavor mixing angles and the lepton
mass ratios;

\item to investigate possible constant
patterns of lepton flavor mixing as the leading-order effects.
\end{enumerate}
The first possibility points to some vanishing (or
vanishingly small) entries of $M^{}_l$ and $M^{}_\nu$, while the
second possibility indicates some equalities or linear
correlations among the entries of $M^{}_l$ or $M^{}_\nu$. In both
cases the underlying flavor symmetries play a crucial role in
deriving the structures of lepton mass matrices which finally
determine the structure of lepton flavor mixing. Of course, how to
pin down the correct flavor symmetries remains an open question.

\subsection{Five Typical Patterns of $U^{}_0$}

It is well known that the special textures of $M^{}_l$ and
$M^{}_\nu$ like that in Eq. (18) can easily be derived from certain
discrete flavor symmetries (e.g., $A^{}_4$ or $S^{}_4$)
\cite{Flavor1,Flavor2}. That is why Eq. (17) formally summarizes a
large class of lepton flavor mixing patterns in which the leading
terms are constant matrices originating from some underlying flavor
symmetries. The fact that $\theta^{}_{13}$ is not very small poses a
meaningful question to us today: can this mixing angle naturally be
generated from the perturbation matrix $\Delta U$? The answer to
this question is certainly dependent upon the form of $U^{}_0$ in
the flavor symmetry limit. Here we reexamine five typical patterns
of $U^{}_0$ in order to get a feeling of the respective structures
of $\Delta U$ which can be constrained by current experimental data
on neutrino oscillations.

For the sake of simplicity, we typically take $\theta^{}_{12} \simeq
34^\circ$, $\theta^{}_{13} \simeq 9^\circ$ and $\theta^{}_{23}
\simeq 45^\circ$ as our inputs to fix the primary structure of the
MNSP matrix $U$. Then we have
\begin{eqnarray}
U = \left( \begin{matrix} 0.819 & 0.552 & 0.156 e^{-i\delta} \cr
-0.395 - 0.092 e^{i\delta} &
0.586 - 0.062 e^{i\delta} &
0.698 \cr 0.395 - 0.092 e^{i\delta} & -0.586 - 0.062
e^{i\delta} & 0.698 \cr \end{matrix} \right) P^{}_\nu \; .
%     (19)
\end{eqnarray}
It makes sense to compare a constant mixing pattern $U^{}_0$ with
the observed pattern of $U$ in Eq. (19), such that one may estimate
the structure of the corresponding perturbation matrix $\Delta U$.
Let us consider five well-known patterns of $U^{}_0$ in the
following for illustration.

(1) The democratic mixing pattern of lepton flavors \cite{FX00,FX96,FX98}:
\begin{eqnarray}
U^{}_0 = \left( \begin{matrix} \frac{1}{\sqrt 2} & \frac{1}{\sqrt 2}
& 0 \cr -\frac{1}{\sqrt 6} & \frac{1}{\sqrt 6} & \frac{\sqrt
2}{\sqrt 3} \cr \frac{1}{\sqrt 3} & -\frac{1}{\sqrt 3} &
\frac{1}{\sqrt 3} \cr \end{matrix} \right) \; ,
%     (20)
\end{eqnarray}
whose three mixing angles are $\theta^{(0)}_{12} = 45^\circ$,
$\theta^{(0)}_{13} = 0^\circ$ and $\theta^{(0)}_{23}
=\arctan(\sqrt{2}) \simeq 54.7^\circ$ in the standard
parametrization as given in Eq. (2). With the help of Eq. (19), we
immediately obtain the form of $\Delta U = UP^\dagger_\nu - U^{}_0$
as follows:
\begin{eqnarray}
\Delta U = \left( \begin{matrix} 0.112
& -0.155 & 0.156 e^{-i\delta} \cr
0.013 - 0.092 e^{i\delta} &
0.178 - 0.062 e^{i\delta} &
-0.118 \cr -0.182 - 0.092 e^{i\delta} & -0.009 - 0.062
e^{i\delta} & 0.121 \cr \end{matrix} \right) \; .
%     (21)
\end{eqnarray}
One can see that the magnitude of each matrix element of $\Delta U$
is of ${\cal O}(0.1)$, implying that the realistic pattern of $U$
might result from a democratic perturbation to $U^{}_0$ (i.e., the
nine entries of $\Delta U$ are all proportional to a common small
parameter).

(2) The bimaximal mixing pattern of lepton flavors \cite{BM1,BM2}:
\begin{eqnarray}
U^{}_0 = \left( \begin{matrix} \frac{1}{\sqrt 2} & \frac{1}{\sqrt 2} & 0
\cr -\frac{1}{2} & \frac{1}{2} & \frac{1}{\sqrt 2} \cr \frac{1}{2} &
-\frac{1}{2} & \frac{1}{\sqrt 2} \cr \end{matrix} \right) \; ,
%     (22)
\end{eqnarray}
which has $\theta^{(0)}_{12} = 45^\circ$, $\theta^{(0)}_{13} =
0^\circ$ and $\theta^{(0)}_{23} = 45^\circ$ in the standard
parametrization. Comparing Eq. (22) with Eq. (19), we obtain the
perturbation matrix
\begin{eqnarray}
\Delta U = \left( \begin{matrix} 0.112
& -0.155 & 0.156 e^{-i\delta} \cr
0.105 - 0.092 e^{i\delta} &
0.086 - 0.062 e^{i\delta} &
-0.009 \cr -0.105 - 0.092 e^{i\delta} & -0.086 - 0.062
e^{i\delta} & -0.009 \cr \end{matrix} \right) \; .
%     (23)
\end{eqnarray}
We see that the matrix elements $(\Delta U)^{}_{\mu 3}$ and $(\Delta
U)^{}_{\tau 3}$ are highly suppressed. In other words, the initially
maximal angle $\theta^{(0)}_{23}$ receives the minimal correction,
which is much smaller than the one received by the initially minimal
angle $\theta^{(0)}_{13}$. Such a situation is more or less
unnatural, at least from a point of view of model building.

(3) The tri-bimaximal mixing pattern of lepton flavors
\cite{TB1,TB2,TB3,TB4}:
\begin{eqnarray}
U^{}_0 = \left( \begin{matrix}  \frac{\sqrt 2}{\sqrt 3} & \frac{1}{\sqrt 3}
& 0 \cr -\frac{1}{\sqrt 6} & \frac{1}{\sqrt 3} & \frac{1}{\sqrt 2}
\cr \frac{1}{\sqrt 6} & -\frac{1}{\sqrt 3} & \frac{1}{\sqrt 2} \cr \end{matrix}
\right) \; ,
%     (24)
\end{eqnarray}
whose three mixing angles are $\theta^{(0)}_{12} =
\arctan(1/\sqrt{2}) \simeq 35.3^\circ$, $\theta^{(0)}_{13} =
0^\circ$ and $\theta^{(0)}_{23} =45^\circ$ in the standard
parametrization. In a similar way we get the corresponding
perturbation matrix
\begin{eqnarray}
\Delta U = \left( \begin{matrix}  0.003
& -0.025 & 0.156 e^{-i\delta} \cr
0.013 - 0.092 e^{i\delta} &
0.009 - 0.062 e^{i\delta} &
-0.009 \cr -0.013 - 0.092 e^{i\delta} & -0.009 - 0.062
e^{i\delta} & -0.009 \cr \end{matrix} \right) \; .
%     (25)
\end{eqnarray}
It is quite obvious that $(\Delta U)^{}_{e 1}$, $(\Delta U)^{}_{e
2}$, $(\Delta U)^{}_{\mu 3}$ and $(\Delta U)^{}_{\tau 3}$ are highly
suppressed. So two initially large angles $\theta^{(0)}_{12}$ and
$\theta^{(0)}_{23}$ are only slightly modified by the perturbation
effects, but the initially minimal angle $\theta^{(0)}_{13}$
receives the maximal correction.

(4) The golden-ratio mixing pattern of lepton flavors
\cite{GR1,GR2}:
\begin{eqnarray}
U^{}_0 = \left( \begin{matrix}  \frac{\sqrt 2}{\sqrt{5 - \sqrt 5}} &
\frac{\sqrt 2}{\sqrt{5 + \sqrt 5}} & 0 \cr -\frac{1}{\sqrt{5 + \sqrt
5}} & \frac{1}{\sqrt{5 - \sqrt 5}} & \frac{1}{\sqrt 2} \cr
\frac{1}{\sqrt{5 + \sqrt 5}} & -\frac{1}{\sqrt{5 - \sqrt 5}} &
\frac{1}{\sqrt 2} \cr \end{matrix} \right) \; ,
%     (26)
\end{eqnarray}
which has $\theta^{(0)}_{12} = \arctan[2/(1+ \sqrt{5})] \simeq
31.7^\circ$, $\theta^{(0)}_{13} = 0^\circ$ and $\theta^{(0)}_{23}
=45^\circ$ in the standard parametrization. In this case the
perturbation matrix $\Delta U$ turns out to be
\begin{eqnarray}
\Delta U = \left( \begin{matrix}  -0.032
& 0.026 & 0.156 e^{-i\delta} \cr
-0.023 - 0.092 e^{i\delta} &
-0.016 - 0.062 e^{i\delta} &
-0.009 \cr 0.023 - 0.092 e^{i\delta} & 0.016 - 0.062
e^{i\delta} & -0.009 \cr \end{matrix} \right) \; .
%     (27)
\end{eqnarray}
Similar to the tri-bimaximal mixing pattern, two initially large
angles of the golden-ratio mixing pattern are only slightly
corrected, but the initially minimal angle $\theta^{(0)}_{13}$ is
significantly modified by the same perturbation.

(5) The hexagonal mixing pattern of lepton flavors
\cite{hexagonal1,hexagonal2,hexagonal3}:
\begin{eqnarray}
U^{}_0 = \left( \begin{matrix}  \frac{\sqrt 3}{2} & \frac{1}{2} & 0
\cr -\frac{\sqrt 2}{4} & \frac{\sqrt 6}{4} & \frac{1}{\sqrt 2} \cr
\frac{\sqrt 2}{4} & -\frac{\sqrt 6}{4} & \frac{1}{\sqrt 2} \cr
\end{matrix} \right) \; ,
%     (28)
\end{eqnarray}
whose mixing angles are $\theta^{(0)}_{12} = 30^\circ$,
$\theta^{(0)}_{13} = 0^\circ$ and $\theta^{(0)}_{23} =45^\circ$ in
the standard parametrization. In this case we obtain the
perturbation matrix
\begin{eqnarray}
\Delta U = \left( \begin{matrix}  -0.047 & 0.052 & 0.156
e^{-i\delta} \cr -0.041 - 0.092 e^{i\delta} & -0.026 - 0.062
e^{i\delta} & -0.009 \cr 0.041 - 0.092 e^{i\delta} & 0.026 - 0.062
e^{i\delta} & -0.009 \cr \end{matrix} \right) \; .
%     (29)
\end{eqnarray}
This result is quite analogous to the one obtained in Eq. (25) or
Eq. (27), simply because the patterns of $U^{}_0$ in these three
cases are quite similar.

Now let us summarize some useful lessons that we can directly learn
from the above five typical examples of $U$.
\begin{itemize}
\item     To accommodate the observed value of
$\theta^{}_{13}$ in a generic flavor mixing structure $U =
\left(U^{}_0 + \Delta U\right) P^{}_\nu$, one has to choose a proper
constant mixing pattern $U^{}_0$ and adjust its perturbation matrix
$\Delta U$. The phenomenological criterion to do so is two-fold: on
the one hand, $U^{}_0$ should easily be derived from a certain
flavor symmetry; on the other hand, $\Delta U$ should have a natural
structure which can easily be accounted for by either the flavor
symmetry breaking or quantum corrections (or both of them).

\item     The common feature of the above five patterns of
$U^{}_0$ is apparently $(U^{}_0)^{}_{e 3} = 0$ (or equivalently,
$\theta^{(0)}_{13} =0^\circ$), implying that a relatively large
perturbation is required for generating $\theta^{}_{13} \sim
9^\circ$. In this case, the closer $\theta^{(0)}_{12}$ and
$\theta^{(0)}_{23}$ are to the observed values of $\theta^{}_{12}$
and $\theta^{}_{23}$, the more unnatural the structure of $\Delta U$
seems to be. The tri-bimaximal mixing pattern given in Eq. (24),
which is currently the most popular ansatz for model building based
on certain flavor symmetries, suffers from this unnaturalness in
particular \cite{correlative}. In this sense we argue that the democratic
mixing pattern in Eq. (29) might be more natural and deserve some
more attention.

\item    One may certainly consider some possible patterns
of $U^{}_0$ which can predict a finite value of $\theta^{(0)}_{13}$
in the vicinity of the experimental value of $\theta^{}_{13}$. In
this case the three mixing angles of $U^{}_0$ may receive comparably
small corrections from the perturbation matrix $\Delta U$, and thus
the naturalness criterion can be satisfied. For example, the
following two patterns of $U^{}_0$ belong to this category and have
been discussed in the literature \cite{RZZ,Feruglio}.
One of them is the so-called correlative mixing pattern
\cite{correlative}
\begin{eqnarray}
U^{}_0 = \left( \begin{matrix}
\frac{\sqrt 2}{\sqrt 3} c^{}_* & \frac{1}{\sqrt 3} c^{}_* &
s^{}_* e^{-i\delta} \cr
-\frac{1}{\sqrt 6} - \frac{1}{\sqrt 3} s^{}_* e^{i\delta} &
\frac{1}{\sqrt 3} - \frac{1}{\sqrt 6} s^{}_* e^{i\delta} &
\frac{1}{\sqrt 2} c^{}_* \cr
\frac{1}{\sqrt 6} - \frac{1}{\sqrt 3} s^{}_* e^{i\delta} &
-\frac{1}{\sqrt 3} - \frac{1}{\sqrt 6} s^{}_* e^{i\delta} &
\frac{1}{\sqrt 2} c^{}_* \cr \end{matrix} \right) \;
%     (30)
\end{eqnarray}
with $c^{}_* \equiv \cos\theta^{}_* = (\sqrt{2} +1)/\sqrt{6}$ and
$s^{}_* \equiv \sin\theta^{}_* = (\sqrt{2} -1)/\sqrt{6}$, which
predicts $\theta^{(0)}_{12} = \arctan(1/\sqrt{2}) \simeq
35.3^\circ$, $\theta^{(0)}_{23} = 45^\circ$ and $\theta^{(0)}_{13} =
\theta^{(0)}_{23} - \theta^{(0)}_{12} \simeq 9.7^\circ$.
The three mixing angles in this constant scenario satisfy
two interesting sum rules:
\begin{eqnarray}
&& \theta^{}_{12} + \theta^{}_{13} = \theta^{}_{23} \; , \nonumber \\
&& \theta^{}_{12} + \theta^{}_{13} +
\theta^{}_{23} = 90^\circ \; ,
%     (31)
\end{eqnarray}
which are geometrically illustrated in Fig. 3.
%%%%%%%%%%%%%%%%%%%%%%%% Fig. 3 %%%%%%%%%%%%%%%%%%%%%%%%
\begin{figure}
\centering
\includegraphics[bb = 230 490 380 710,scale=0.6]{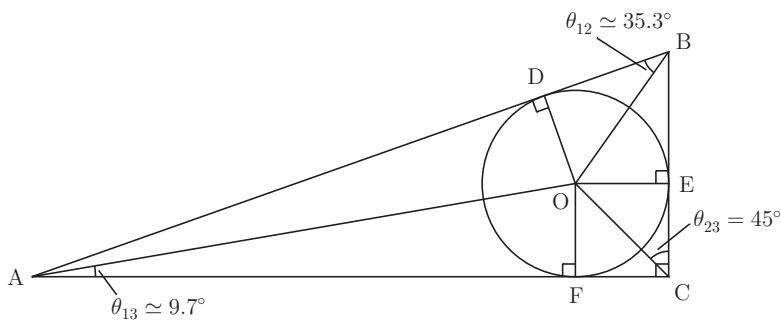}
\caption{A geometrical description of the sum rules $\theta^{}_{12}
+ \theta^{}_{13} = \theta^{}_{23}$ and $\theta^{}_{12} +
\theta^{}_{13} + \theta^{}_{23} = 90^\circ$ for the correlative
neutrino mixing pattern in terms of the inner angles of the right
triangle $\triangle\rm ABC$.}
\end{figure}
%%%%%%%%%%%%%%%%%%%%%%%%%%%%%%%%%%%%%%%%%%%%%%%%
The other pattern of $U^{}_0$ is the tetra-maximal mixing pattern
\cite{Xing08}
\begin{eqnarray}
U^{}_0 = \left( \begin{matrix}  \frac{2 + \sqrt 2}{4} &
\frac{1}{2} & \frac{2 - \sqrt 2}{4} \cr
-\frac{\sqrt{2}}{4} + \frac{i \left(\sqrt{2} - 1\right)}{4} &
\frac{1}{2} - \frac{i\sqrt{2}}{4} &
\frac{\sqrt{2}}{4} + \frac{i \left(\sqrt{2} + 1\right)}{4} \cr
-\frac{\sqrt{2}}{4} - \frac{i \left(\sqrt{2} - 1\right)}{4} &
\frac{1}{2} + \frac{i\sqrt{2}}{4} & \frac{\sqrt{2}}{4} -
\frac{i \left(\sqrt{2} + 1 \right)}{4} \cr \end{matrix} \right) \; ,
%     (32)
\end{eqnarray}
which predicts $\theta^{(0)}_{12} = \arctan(2-\sqrt{2})\simeq
30.4^\circ$, $\theta^{(0)}_{23} = 45^\circ$ and $\theta^{(0)}_{13} =
\arcsin[(2-\sqrt{2})/4] \simeq 8.4^\circ$. Of course, whether such
constant mixing patterns can easily be derived from some underlying
flavor symmetries remains an open question.
\end{itemize}
In short, today's model building has to take the challenge caused by
the reasonably large value of $\theta^{}_{13}$.

Note that the RGE running effects or finite quantum corrections are not easy
to generate $\theta^{}_{13} \simeq 9^\circ$ from $\theta^{(0)}_{13}
= 0^\circ$, unless the seesaw threshold effects or other extreme
conditions are taken into account
\cite{RGE01,RGE02,RGE03,RGE04,RGE05,RGE06,RGE07,RGE08,RGE09,RGE10}.
One may therefore consider a pattern of $U^{}_0$ with nonzero
$\theta^{(0)}_{13}$, such as the tetra-maximal mixing pattern
\cite{ZZ} or the correlative mixing pattern \cite{LX12}, as a
starting point of view to calculate the radiative corrections before
confronting it with current experimental data. We shall elaborate on
this point in detail in section 5.

\subsection{The Minimal Perturbation to $U^{}_0$}

Note that the perturbation matrix $\Delta U$ in Eq. (17) is in
general a sum of all possible perturbations to the constant flavor
mixing matrix $U^{}_0$. From the point of view of model building, it
is helpful to single out a viable $\Delta U$ whose form is as simple
as possible. To do so, let us reexpress Eq. (17) in the following
manner:
\begin{eqnarray}
U = \left(U^{}_0 + \Delta U\right) P^{}_\nu = U^{}_0 \left({\bf 1} +
\Delta U^\prime \right) P^{}_\nu = \left({\bf 1} + \Delta
U^\prime_{\rm L} \right) U^{}_0 \left({\bf 1} + \Delta U^\prime_{\rm
R} \right) P^{}_\nu  \; ,
%     (33)
\end{eqnarray}
where $\Delta U = U^{}_0 \Delta U^\prime = \Delta U^\prime_{\rm L}
U^{}_0 + U^{}_0 \Delta U^\prime_{\rm R} + \Delta U^\prime_{\rm L}
U^{}_0 \Delta U^\prime_{\rm R}$ holds, and it satisfies the
condition $U^{}_0 \Delta U^\dagger + \Delta U U^\dagger_0 + \Delta U
\Delta U^\dagger = {\bf 0}$ as a result of the unitarity of $U$
itself. Therefore, one may achieve a viable but minimal perturbation
to $U^{}_0$ by switching off $\Delta U^\prime_{\rm L}$ (or $\Delta
U^\prime_{\rm R}$) and adjusting $\Delta U^\prime_{\rm R}$ (or
$\Delta U^\prime_{\rm L}$) to its simplest form which is allowed by
current experimental data. Such a treatment is actually equivalent
to multiplying $U^{}_0$ by a unitary perturbation matrix, which may
more or less deviate from the unit matrix $\bf 1$, from either
its left-hand side or its right-hand side. The first example of this
kind was given before \cite{FX00,FX96,FX98} for the democratic
mixing pattern, and its $\Delta U$ was mainly responsible for the
generation of nonzero $\theta^{}_{13}$ and $\delta$.

Here we concentrate on the typical patterns of $U^{}_0$ discussed
above and outline the main ideas of choosing the minimal
perturbations to them.
\begin{itemize}
\item     If $U^{}_0$ predicts $\theta^{(0)}_{23} =
45^\circ$ and $\theta^{(0)}_{13} =0^\circ$ together with $\theta^{(0)}_{12}
>34^\circ$ (the best-fit value based on current neutrino oscillation
data \cite{Fogli}), then the simplest way to generate a relatively
large $\theta^{}_{13}$, keep $\theta^{}_{23} = \theta^{(0)}_{23} =
45^\circ$ unchanged and correct $\theta^{(0)}_{12}$ to a slightly
smaller value is to choose a complex $(2,3)$ rotation matrix as the
perturbation matrix:
\begin{eqnarray}
{\bf 1} + \Delta U^\prime = \left( \begin{matrix}  1 & 0 & 0 \cr 0 &
\cos\theta & i \sin\theta \cr 0 & i\sin\theta & \cos\theta \cr
\end{matrix} \right) ~~ {\rm or} ~~ \Delta U^\prime \simeq \left(
\begin{matrix}  0 & 0 & 0 \cr 0 & -\frac{1}{2}\sin^2\theta &
i\sin\theta \cr 0 & i\sin\theta & -\frac{1}{2}\sin^2\theta \cr
\end{matrix} \right) \; ,
%     (34)
\end{eqnarray}
where $\theta$ is a small angle to trigger the perturbation effect.
The most striking example in this category is to take $U^{}_0$ to be
the tri-bimaximal mixing pattern given in Eq. (24). The result is
\cite{XZ07,Zhou12}:
\begin{eqnarray}
U = \left ( \begin{matrix} \frac{\sqrt 2}{\sqrt 3} & \frac{1}{\sqrt
3} \cos\theta & \frac{i}{\sqrt 3} \sin\theta \cr -\frac{1}{\sqrt 6}
& \frac{1}{\sqrt 3} \cos\theta + \frac{i}{\sqrt 2} \sin\theta &
\frac{1}{\sqrt 2} \cos\theta + \frac{i}{\sqrt 3} \sin\theta \cr
\frac{1}{\sqrt 6} & -\frac{1}{\sqrt 3} \cos\theta + \frac{i}{\sqrt
2} \sin\theta & \frac{1}{\sqrt 2} \cos\theta - \frac{i}{\sqrt 3}
\sin\theta \cr \end{matrix} \right ) P^{}_\nu \; ,
%     (35)
\end{eqnarray}
which predicts
\begin{eqnarray}
\sin^2 \theta^{}_{12} = \frac{1}{3} \left(1 - 2 \tan^2
\theta^{}_{13} \right) \; , ~ \sin^2 \theta^{}_{13} = \frac{1}{3}
\sin^2 \theta \; , ~ \theta^{}_{23} = 45^\circ \; , ~ \delta =
90^\circ \;
%     (36)
\end{eqnarray}
in the standard parametrization. Note that the obtained correlation
between $\theta^{}_{12}$ and $\theta^{}_{13}$ is especially
interesting because it leads us to $\theta^{}_{12} \to 34^\circ$
when $\theta^{}_{13} \to 9^\circ$, consistent with the present
experimental data. If $\theta^{}_{23}$ is allowed to slightly
deviate from $\theta^{(0)}_{23} =45^\circ$, then one may simply make
the replacement $i \to e^{i\delta}$ in Eq. (35).

\item     If $U^{}_0$ predicts $\theta^{(0)}_{23} =
45^\circ$ and $\theta^{(0)}_{13} =0^\circ$ together with
$\theta^{(0)}_{12} <34^\circ$, then the most economical way to
generate a relatively large $\theta^{}_{13}$, keep $\theta^{}_{23} =
\theta^{(0)}_{23} = 45^\circ$ unchanged and correct
$\theta^{(0)}_{12}$ to a slightly larger value is to choose a
complex $(1,3)$ rotation matrix as the perturbation matrix:
\begin{eqnarray}
{\bf 1} + \Delta U^\prime = \left( \begin{matrix}  \cos\theta & 0 &
i\sin\theta \cr 0 & 1 & 0 \cr i\sin\theta & 0 & \cos\theta \cr
\end{matrix} \right) ~~ {\rm or} ~~ \Delta U^\prime \simeq \left(
\begin{matrix} -\frac{1}{2}\sin^2\theta & 0 & i\sin\theta \cr 0 & 0
& 0 \cr i\sin\theta & 0 & -\frac{1}{2}\sin^2\theta \cr \end{matrix}
\right) \; .
%     (37)
\end{eqnarray}
Taking $U^{}_0$ to be the golden-ratio mixing pattern in Eq. (26),
we immediately arrive at
\begin{eqnarray}
U = \left( \begin{matrix}  \frac{\sqrt 2}{\sqrt{5 - \sqrt 5}}
\cos\theta & \frac{\sqrt 2}{\sqrt{5 + \sqrt 5}}  & \frac{i\sqrt
2}{\sqrt{5 - \sqrt 5}} \sin\theta \cr -\frac{1}{\sqrt{5 + \sqrt 5}}
\cos\theta + \frac{i}{\sqrt 2} \sin\theta & \frac{1}{\sqrt{5 - \sqrt
5}} & \frac{1}{\sqrt 2} \cos\theta - \frac{i}{\sqrt{5 + \sqrt 5}}
\sin\theta \cr \frac{1}{\sqrt{5 + \sqrt 5}} \cos\theta +
\frac{i}{\sqrt 2} \sin\theta & -\frac{1}{\sqrt{5 - \sqrt 5}} &
\frac{1}{\sqrt 2} \cos\theta + \frac{i}{\sqrt{5 + \sqrt 5}}
\sin\theta \cr \end{matrix} \right) P^{}_\nu \; ,
%     (38)
\end{eqnarray}
whose predictions include $\theta^{}_{23} = 45^\circ$,
$\delta = 90^\circ$, and
\begin{eqnarray}
\sin^2\theta^{}_{12} = \frac{2}{5 + \sqrt 5} \left( 1 +
\tan^2\theta^{}_{13} \right) \; , ~~~~~
\sin^2\theta^{}_{13} = \frac{2}{5 - \sqrt 5} \sin^2\theta
%     (39)
\end{eqnarray}
in the standard parametrization of $U$. In this case the correlation
between $\theta^{}_{12}$ and $\theta^{}_{13}$ leads to
$\theta^{}_{12} \to 32^\circ$ when $\theta^{}_{13} \to 9^\circ$,
compatible with the experimental data. Again, the replacement $i
\to e^{i\delta}$ in Eq. (38) allows one to obtain a somewhat more
flexible value of $\theta^{}_{23}$ which may slightly deviate from
$\theta^{(0)}_{23} =45^\circ$.

\item     If $U^{}_0$ is quite far away from the
realistic MNSP matrix $U$, one has to consider a somewhat
complicated perturbation matrix including two rotation angles. In
the neglect of CP violation, for instance, we may consider
\begin{eqnarray}
{\bf 1} + \Delta U^\prime = \left( \begin{matrix}  c^\prime_{12} &
-s^\prime_{12} & 0 \cr s^\prime_{12} c^\prime_{23} & c^\prime_{12}
c^\prime_{23} & s^\prime_{23} \cr s^\prime_{12} s^\prime_{23} &
c^\prime_{12} s^\prime_{23} & -c^\prime_{23} \cr \end{matrix} \right) \; ,
%     (40)
\end{eqnarray}
where $c^\prime_{ij} \equiv \cos\theta^\prime_{ij}$ and
$s^\prime_{ij} \equiv \sin\theta^\prime_{ij}$ (for $ij = 12, 23$).
However, we hope that the resulting structure of $U$ still allows us
to obtain one or two predictions, in particular for the mixing angle
$\theta^{}_{13}$. A simple example of this kind has been given before
\cite{Xing12} by taking $U^{}_0$ to be the democratic mixing
pattern, and it predicts an interesting relationship between
$\theta^{}_{13}$ and $\theta^{}_{23}$ in the standard
parametrization:
\begin{eqnarray}
\sin\theta^{}_{13} = \frac{\sqrt{2} - \tan\theta^{}_{23}}
{\sqrt{5 - 2\sqrt{2} \tan\theta^{}_{23} + 4 \tan^2 \theta^{}_{23}}} \; .
%     (41)
\end{eqnarray}
Typically taking $\theta^{}_{23} \simeq 45^\circ$, we can arrive at
$\theta^{}_{13} \simeq 9.6^\circ$ \cite{Xing12}. It is certainly easy to
accommodate a CP-violating phase in $\Delta U^\prime$,
although its form might not be really minimal anymore.
\end{itemize}
For those constant flavor mixing patterns with $\theta^{(0)}_{13}
\neq 0^\circ$ from the very beginning, such as the correlative
\cite{correlative} and tetra-maximal \cite{Xing08} mixing scenarios
given in Eqs. (30) and (32), the similar minimal perturbations can
be introduced in order to make the resulting MNSP matrix $U$ fit the
experimental data to a better degree of accuracy.

It should be noted that the above discussions about possible
patterns of $\Delta U$ (or $\Delta U^\prime$) with respect to those
of $U^{}_0$ are purely phenomenological. From the point of view of
model building, it is more meaningful to consider the textures of
lepton mass matrices
\begin{eqnarray}
M^{}_l = M^{(0)}_l + \Delta M^{}_l \; , ~~~~~~~~~~~~
M^{}_\nu = M^{(0)}_\nu + \Delta M^{}_\nu \; ,
%     (42)
\end{eqnarray}
where $M^{(0)}_l$ and $M^{(0)}_\nu$ can be obtained in the limit of
certain flavor symmetries, and their special structures allow us to
achieve a constant flavor mixing pattern $U^{}_0$. The perturbation
matrices $\Delta M^{}_l$ and $\Delta M^{}_\nu$ play an important
role in transforming $U^{}_0$ into the realistic MNSP matrix $U$,
and thus their textures should be determined in a simple way and
with a good reason. The connection between $\Delta M^{}_{l,\nu}$ and
$\Delta U$ (or $\Delta U^\prime$) depends on the details of a lepton
flavor model and may not be very transparent in most cases. In the
basis where $M^{}_l$ is real and positive, however, $\Delta
M^{}_\nu$ can be formally expressed as
\begin{eqnarray}
\Delta M^{}_\nu = \left(U^{}_0 + \Delta U\right) \overline{M}^{}_\nu
\left(U^{}_0 + \Delta U\right)^T - U^{}_0 \overline{M}^{(0)}_\nu
U^{T}_0 \; ,
%     (43}
\end{eqnarray}
where $\overline{M}^{}_\nu = P^{}_\nu \widehat{M}^{}_\nu P^T_\nu$
and $\overline{M}^{(0)}_\nu = P^{\prime}_\nu
\widehat{M}^{\prime}_\nu P^{\prime T}_\nu$,
$\widehat{M}^{\prime}_\nu \equiv {\rm Diag}\{m^{\prime}_1,
m^{\prime}_2, m^{\prime}_3\}$ and $P^\prime_\nu \equiv {\rm
Diag}\{e^{i\rho^\prime}, e^{i\sigma^\prime}, 1\}$. Here
$m^{\prime}_i$ (for $i=1,2,3$) denote the eigenvalues of
$M^{(0)}_\nu$ in the symmetry limit, while $\rho^\prime$ and
$\sigma^\prime$ stand for the Majorana phases in the same limit. It
is therefore possible, at least in principle, to fix the structure
of $\Delta M^{}_\nu$ with the help of a certain flavor symmetry and
current experimental data.

\section{UNSUPPRESSED $\theta^{}_{13}$ AND RGE RUNNING EFFECTS}

The RGE running effects of the neutrino flavor parameters have been
discussed in many papers
\cite{RGE01,RGE02,RGE03,RGE04,RGE05,RGE06,RGE07,RGE08,RGE09,RGE10,RGE11,RGE12,RGE13,RGE14,RGE15,RGE16,RGE17,RGE18,RGE19,RGE20}. It is known that large radiative corrections to those
parameters are possible, especially when the neutrino masses
are nearly degenerate or the value of $\tan\beta$ is sufficiently
big in the MSSM. The fact that
$\theta^{}_{13}$ is not as small as previously expected motivates one
to reconsider how it can be generated at the tree level or by
quantum corrections. Some studies in this regard have recently been
done to look at the impacts of a relatively large $\theta^{}_{13}$ on
the running behaviors of the other two mixing angles and the
CP-violating phases \cite{LX12}.

\subsection{Approximate RGEs}

The masses of the Majorana neutrinos are believed to be attributed
to some underlying new physics at a superhigh-energy scale $\Lambda$
(e.g., via the canonical seesaw mechanism
\cite{SS1,SS2,SS3,SS4,SS5}). But this kind of new physics can all
point to the unique dimension-5 Weinberg operator for the neutrino
masses in an effective field theory after the corresponding heavy
degrees of freedom are integrated out \cite{Weinberg1979}. In the
MSSM, such a dimension-5 operator reads
\begin{eqnarray}
\frac{{\cal L}^{}_{\rm d=5}}{\Lambda} & = & \frac{1}{2} \;
\overline{\ell^{}_{\rm L}} H^{}_{2} \cdot \kappa \cdot H^{T}_{2}
\ell^{c}_{\rm L} \; + \; {\rm h.c.} \; ,
%       (44)
\end{eqnarray}
where $\Lambda$ denotes the cutoff scale, $\ell^{}_{\rm L}$ stands
for the left-handed lepton doublet, $H^{}_2$ is one of the MSSM
Higgs doublets, and $\kappa$ represents the effective neutrino
coupling matrix. One may obtain the effective Majorana neutrino mass
matrix $M^{}_{\nu} = \kappa v^2 \tan^2\beta/(1 + \tan^2\beta)$ after
spontaneous gauge symmetry breaking. The cutoff scale $\Lambda$
implies the scale of new physics, such as the mass scale
of the heavy Majorana neutrinos in the canonical seesaw mechanism.
The evolution of $\kappa$ from $\Lambda$ down to the electroweak
scale $\Lambda^{}_{\rm EW}$ is formally independent of any details
of the relevent model from which $\kappa$ is derived. Below
$\Lambda$ the scale dependence of $\kappa$ is described by
\begin{equation}
16\pi^2 \frac{{\rm d}\kappa}{{\rm d}t} \; = \; \alpha^{}_{\rm M}
\kappa + \left [ \left ( Y^{}_{l} Y^{\dagger}_{l} \right ) \kappa +
\kappa \left ( Y^{}_{l} Y^{\dagger}_{l} \right )^{T}_{} \right ] \;
%       (45)
\end{equation}
at the one-loop level in the MSSM \cite{RGE11,RGE12,RGE14}, where
$\alpha^{}_{\rm M} \approx - 1.2 g^{2}_{1} - 6 g^{2}_{2} + 6
y^{2}_{t}$.

One may use Eq. (45) to derive the explicit RGEs of the three
neutrino masses and six flavor mixing parameters
\cite{RGE04,RGE05,RGE11,RGE12,RGE14,RGE17}. Given an approximate
mass degeneracy of the three neutrinos together with the standard
parametrization of the MNSP matrix $U$ in Eq. (2),
the RGEs of $m^{}_i$ (for $i=1,2,3$) turn out to be
\begin{eqnarray}
\frac{{\rm d} m^{}_1}{{\rm d} t} & \approx & \frac{m^{}_1}{16\pi^2}
\left [ \alpha^{}_{\rm M}  + 2 y^{2}_{\tau} \left (s^{2}_{12}
s^{2}_{23} - 2 c^{}_{\delta} c^{}_{12} c^{}_{23} s^{}_{12} s^{}_{23}
s^{}_{13}  + {\cal O} (s^{2}_{13}) \right ) \right ] \; ,
\nonumber \\
\frac{{\rm d} m^{}_2}{{\rm d} t} & \approx & \frac{m^{}_2}{16\pi^2}
\left [ \alpha^{}_{\rm M} + 2 y^{2}_{\tau} \left ( c^{2}_{12}
s^{2}_{23} + 2 c^{}_{\delta } c^{}_{12} c^{}_{23} s^{}_{12}
s^{}_{23} s^{}_{13}  + {\cal O} (s^{2}_{13}) \right ) \right ] \; ,
\nonumber \\
\frac{{\rm d} m^{}_3}{{\rm d} t} & \approx & \frac{m^{}_3}{16\pi^2}
\left [ \alpha^{}_{\rm M}  + 2 y^{2}_{\tau} c^{2}_{23} + {\cal O}
(s^{2}_{13}) \right ] \; .
%       (46)
\end{eqnarray}
The RGEs of $\theta^{}_{ij}$ (for $ij =12, 23, 13$) are found to be
\begin{eqnarray}
\frac{{\rm d} \theta^{}_{12}}{{\rm d} t} & \approx & -
\frac{y^{2}_{\tau}}{4 \pi^2} \left \{ \frac{m^{2}_{1}}{\Delta
m^{2}_{21}} s^{}_{23} \left [ \left ( c^{}_{12} s^{}_{12} s^{}_{23} -
\cos2\theta^{}_{12} c^{}_{23} s^{}_{13} c^{}_{\delta}
\right ) c^{}_{(\rho - \sigma)} \right. \right. \nonumber\\
& & ~~~~~~~~~~~~~~~~~~~~~ \left. \left. + c^{}_{23} s^{}_{13}
s^{}_{\delta} s^{}_{(\rho - \sigma)} \right ] c^{}_{(\rho - \sigma)}
\right.
\nonumber\\
& & ~~~ \left. - \frac{m^{2}_{1}}{\Delta m^{2}_{32}} c^{}_{23}
s^{}_{23} s^{}_{13} \left ( s^{2}_{12} c^{}_{(\delta + \rho)}
c^{}_{\rho} + c^{2}_{12} c^{}_{(\delta + \sigma)} c^{}_{\sigma}
\right )  + {\cal O} (s^{2}_{13}) \right \} \; ,
\nonumber \\
\nonumber \\
\frac{{\rm d} \theta^{}_{23}}{{\rm d} t} & \approx & -
\frac{y^{2}_{\tau}}{4 \pi^2} \frac{m^{2}_{1}}{\Delta m^{2}_{32}}
c^{}_{23} \left [s^{}_{23} \left ( s^{2}_{12} c^{2}_{\rho}
+c^{2}_{12} c^{2}_{\sigma} \right ) \right. \nonumber\\
& & ~~~  \left. - \frac{1}{2} c^{}_{12}
s^{}_{12} c^{}_{23} s^{}_{13} \left ( c^{}_{(\delta + 2 \rho)} -
c^{}_{(\delta + 2 \sigma)} \right ) + {\cal O} (s^{2}_{13})\right ]
\; ,
\nonumber \\
\nonumber \\
\frac{{\rm d} \theta^{}_{13}}{{\rm d} t} & \approx &
\frac{y^{2}_{\tau}}{8 \pi^2} \frac{m^{2}_{1}}{\Delta m^{2}_{32}}
c^{}_{23} c^{}_{13} \left [ c^{}_{12} s^{}_{12} s^{}_{23} \left (
c^{}_{(\delta + 2 \rho)} - c^{}_{(\delta + 2 \sigma)} \right ) \right.
\nonumber\\[2mm]
& & \left. ~~~ - 2
c^{}_{23} s^{}_{13} \left ( c^{2}_{12} c^{2}_{(\delta + \rho)} +
s^{2}_{12} c^{2}_{(\delta + \sigma)} \right ) + {\cal O}
(s^{2}_{13}) \right ] \; ,
%       (47)
\end{eqnarray}
in which $c^{}_{x} \equiv \cos x$ and $s^{}_{x} \equiv \sin x$ (for
$x = \delta, \; \rho, \; \sigma, \; \rho-\sigma, \; \delta+\rho, \;
\delta+\sigma, \; \delta+2\rho, \; \delta+2\sigma$). The RGEs of the
three CP-violating phases $\delta$, $\rho$ and $\sigma$ can be
written as
\begin{eqnarray}
\frac{{\rm d} \delta}{{\rm d} t} & \approx & - \frac{y^{2}_{\tau}}{4
\pi^2} \left \{ \frac{m^{2}_{1}}{\Delta m^{2}_{21}} s^{}_{23} \left
[ \left ( s^{}_{23} - \frac{\cos2\theta^{}_{12} c^{}_{23} s^{}_{13}
c^{}_{\delta}}{c^{}_{12} s^{}_{12}} \right ) c^{}_{(\rho - \sigma)}
\right. \right. \nonumber\\
& & ~~~~~~~~~~~~~~~ \left. \left. + \frac{c^{}_{23} s^{}_{13}
s^{}_{\delta}}{c^{}_{12} s^{}_{12}} s^{}_{(\rho - \sigma)} + {\cal
O} (s^{2}_{13}) \right ] s^{}_{(\rho - \sigma)} \right.
\nonumber\\
& & ~~~ \left. - \frac{m^{2}_{1}}{\Delta m^{2}_{32}}
s^{-1}_{13} \left [ \frac{1}{2} c^{}_{12} s^{}_{12} c^{}_{23}
s^{}_{23} \left ( s^{}_{(\delta + 2 \rho)} - s^{}_{(\delta +
2 \sigma)} \right )
\right. \right. \nonumber\\[2mm]
& & ~~~~~~~~~~~~ \left. \left. + \left ( c^{}_{(\delta -
\rho)} s^{}_{(\delta - \rho)} s^{2}_{12} +c^{}_{(\delta - \sigma)}
s^{}_{(\delta - \sigma)} c^{2}_{12} \right ) \cos2\theta^{}_{23}
s^{}_{13} \right. \right. \nonumber\\
&& ~~~~~~~~~~~~ \left. \left. + \left ( c^{}_{\rho} s^{}_{\rho}
c^{2}_{12} + c^{}_{\sigma} s^{}_{\sigma} s^{2}_{12} \right )
c^{2}_{23} s^{}_{13} + {\cal O} (s^{2}_{13}) \right ] \right \} \; ,
\nonumber \\
\nonumber \\
\frac{{\rm d} \rho}{{\rm d} t} & \approx & \frac{y^{2}_{\tau}}{4
\pi^2} \left \{ \frac{m^{2}_{1}}{\Delta m^{2}_{21}} s^{}_{23}
c^{2}_{12} \left [ \left ( s^{}_{23} - \frac{\cos2\theta^{}_{12}
c^{}_{23} s^{}_{13} c^{}_{\delta}}{c^{}_{12} s^{}_{12}} \right )
c^{}_{(\rho - \sigma)} \right. \right. \nonumber\\
& & ~~~~~~~~~~~~~~~ \left. \left. + \frac{c^{}_{23} s^{}_{13}
s^{}_{\delta}}{c^{}_{12} s^{}_{12}} s^{}_{(\rho - \sigma)} + {\cal
O} (s^{2}_{13}) \right ] s^{}_{(\rho - \sigma)} \right.
\nonumber\\
& & \left. - \frac{m^{2}_{1}}{\Delta m^{2}_{32}} s^{-1}_{13}
\left [ \left ( c^{}_{(\delta - \rho)} s^{}_{(\delta - \rho)}
s^{2}_{12} +c^{}_{(\delta - \sigma)} s^{}_{(\delta - \sigma)}
c^{2}_{12} \right ) \cos2\theta^{}_{23} s^{}_{13} + {\cal O}
(s^{2}_{13}) \right ] \right \} \; ,
\nonumber \\
\nonumber \\
\frac{{\rm d} \sigma}{{\rm d} t} & \approx & \frac{y^{2}_{\tau}}{4
\pi^2} \left \{ \frac{m^{2}_{1}}{\Delta m^{2}_{21}} s^{}_{23}
s^{2}_{12} \left [ \left (s^{}_{23} - \frac{\cos2\theta^{}_{12}
c^{}_{23} s^{}_{13} c^{}_{\delta}}{c^{}_{12} s^{}_{12}} \right )
c^{}_{(\rho - \sigma)} \right. \right. \nonumber\\
& & ~~~~~~~~~~~~~~~ \left. \left. + \frac{c^{}_{23} s^{}_{13}
s^{}_{\delta}}{c^{}_{12} s^{}_{12}} s^{}_{(\rho - \sigma)} + {\cal
O} (s^{2}_{13}) \right ] s^{}_{(\rho - \sigma)} \right.
\nonumber\\
& & \left. - \frac{m^{2}_{1}}{\Delta m^{2}_{32}} s^{-1}_{13}
\left [ \left ( c^{}_{(\delta - \rho)} s^{}_{(\delta - \rho)}
s^{2}_{12} +c^{}_{(\delta - \sigma)} s^{}_{(\delta - \sigma)}
c^{2}_{12} \right ) \cos2\theta^{}_{23} s^{}_{13} + {\cal O}
(s^{2}_{13}) \right ] \right \} \; . \nonumber\\
%       (48)
\end{eqnarray}
In addition, the RGE of ${\cal J}$ is obtained as follows:
\begin{eqnarray}
\frac{\rm d}{{\rm d}t} \; {\cal J} & \approx & -
\frac{y^2_\tau}{8\pi^2} \left \{ \frac{m^{2}_{1}}{\Delta m^{2}_{21}} \left
[ {\cal J} \cos2\theta^{}_{12} s^{2}_{23} -
\cos^{2}_{}2\theta^{}_{12} c^{2}_{23} s^{2}_{23} c^{2}_{13}
s^{2}_{13} c^{}_{\delta} s^{}_{\delta} \right ] \right. \nonumber\\
& & ~~~~~~ \left. + \; \frac{m^{2}_{1}}{\Delta m^{2}_{32}} {\cal
J} \cos2\theta^{}_{23} + {\cal O} (s^{3}_{13}) \right \} \; .
%       (49)
\end{eqnarray}
The running behaviors of the three mixing angles and three
CP-violating phases for a very small $\theta^{}_{13}$ can
be very different from those for a relative large $\theta^{}_{13}$. In
particular, the CP-violating phases play a crucial role in the RGEs.
Let us elaborate on this point in the following.

\subsection{Flavor Mixing Angles and CP-violating Phases}

(1) {\it The running behaviors of the three mixing angles}.
Eq. (47) shows that the RGE running behaviors of the three neutrino
mixing angles are strongly dependent on the three CP-violating
phases. As for the Majorana neutrinos, the radiative corrections to
the three mixing angles can be adjusted by choosing different values
of the CP-violating phases $\delta$, $\rho$ and $\sigma$. A
numerical analysis has been carried out to look at their numerical
evolution to $\Lambda^{}_{\rm FS}$ via the RGEs in the MSSM with
$\tan\beta = 10$ (denoted as ``MSSM10'' for short) or
$\tan\beta = 50$ (denoted as ``MSSM50'' for short), in which
$\theta^{}_{12} = 34^\circ$, $\theta^{}_{23} = 46^\circ$,
$\theta^{}_{13} = 9^\circ$ are taken as the typical input and the three
CP-violating phases are freely adjusted at $\Lambda^{}_{\rm EW}$
\cite{LX12}. The main results are summarized in Table 2, where the
upper (lower) lines show the possible ranges of three mixing angles
at $\Lambda^{}_{\rm FS}$ for the normal (inverted) neutrino mass
hierarchy.
%%%%%%%%%%%%%%%%%%%%%%%%%%%%%% Table 2 %%%%%%%%%%%%%%%%%%%%%%%%%%
\begin{table}
\tbl{The radiative corrections to the three mixing angles from
$\Lambda^{}_{\rm EW} \sim 10^{2}$ GeV to $\Lambda^{}_{\rm FS} \sim
10^{14}$ GeV in the MSSM with $\tan\beta =10$ or $50$, where the three
CP-violating phases are freely adjusted.} {\begin{tabular}{lllll}
\hline Parameter ~ & Input ($\Lambda^{}_{\rm EW}$) &
\multicolumn{2}{l}{Output
($\Lambda^{}_{\rm FS}$)} \\
\cline{3-5}
& & MSSM10 & MSSM50 \\
\hline
\multirow{2}{*}{~ $\theta^{}_{12}$} & \multirow{2}{*}{~ $34.0^\circ$} &
$7^\circ$ to $55^\circ$ & $0.5^\circ$ to $62^\circ$ & (NH) \\
& & $2^\circ$ to $31^\circ$ & $2^\circ$ to $45^\circ$ & (IH) \\
\hline
\multirow{2}{*}{~ $\theta^{}_{23}$} & \multirow{2}{*}{~ $46.0^\circ$} &
$47^\circ$ to $48.5^\circ$ & $7.5^\circ$ to $45.5^\circ$ & (NH) \\
& & $43.5^\circ$ to $45^\circ$ & $46.5^\circ$ to $89^\circ$ & (IH) \\
\hline
\multirow{2}{*}{~ $\theta^{}_{13}$} & \multirow{2}{*}{~ $9.0^\circ$} &
$8.5^\circ$ to $10^\circ$ & $2^\circ$ to $23^\circ$ & (NH) \\
& & $7.5^\circ$ to $9.5^\circ$ & $7.5^\circ$ to $82^\circ$ & (IH) \\
\hline
\end{tabular}}
\end{table}
%%%%%%%%%%%%%%%%%%%%%%%%%%%%%%%%%%%%%%%%%%%%%%%%%%%%%%%%%%%%%%%%%%

The RGE of $\theta^{}_{12}$ is dominated by the term
$\displaystyle - \frac{y^{2}_{\tau}}{8 \pi^2}
\frac{m^{2}_{1}}{\Delta m^{2}_{21}} c^{}_{12} s^{}_{12} s^{2}_{23}
c^{2}_{(\rho - \sigma)} $, implying that the magnitude of the
radiative correction to $\theta^{}_{12}$ depends strongly on the
phase difference $(\rho - \sigma)$. Hence $\theta^{}_{12}$ is most
sensitive to the RGE effect when $\rho \simeq \sigma$ holds. Running
from $\Lambda^{}_{\rm FS} \sim 10^{14}_{}$ GeV down to
$\Lambda^{}_{\rm EW}$, the mixing angles $\theta^{}_{23}$ and
$\theta^{}_{13}$ receive less significant radiative corrections in
the MSSM10 case, as shown in Table 2. While in the MSSM50 case
$\theta^{}_{23}$ and $\theta^{}_{13}$ may also receive significant
radiative corrections if three CP-violating phases are well turned.

We have seen that the values of the three CP-violating phases are
crucial for the evolution of the three mixing angles. A very special
case is $(\rho - \sigma) \simeq \pm 90^\circ$, which leads us to
\begin{eqnarray}
\frac{{\rm d} \theta^{}_{12}}{{\rm d} t} & \approx &
\frac{y^{2}_{\tau}}{4 \pi^2}  \frac{m^{2}_{1}}{\Delta m^{2}_{32}}
c^{}_{23} s^{}_{23} s^{}_{13} \left ( s^{2}_{12} c^{}_{(\delta +
\rho)} c^{}_{\rho} + c^{2}_{12} s^{}_{(\delta + \rho)} s^{}_{\rho}
\right ) \; ,
\nonumber \\
\nonumber \\
\frac{{\rm d} \theta^{}_{23}}{{\rm d} t} & \approx & -
\frac{y^{2}_{\tau}}{4 \pi^2} \frac{m^{2}_{1}}{\Delta m^{2}_{32}}
c^{}_{23} \left [ s^{}_{23} \left ( s^{2}_{12} c^{2}_{\rho}
+c^{2}_{12} s^{2}_{\rho} \right ) -  c^{}_{12} s^{}_{12} c^{}_{23}
s^{}_{13} c^{}_{(\delta + 2 \rho)} \right ] \; ,
\nonumber \\
\nonumber \\
\frac{{\rm d} \theta^{}_{13}}{{\rm d} t} & \approx &
\frac{y^{2}_{\tau}}{4 \pi^2} \frac{m^{2}_{1}}{\Delta m^{2}_{32}}
c^{}_{23} c^{}_{13} \left [ c^{}_{12} s^{}_{12} s^{}_{23}
c^{}_{(\delta + 2 \rho)} - c^{}_{23} s^{}_{13} \left ( c^{2}_{12}
c^{2}_{(\delta + \rho)} + s^{2}_{12} s^{2}_{(\delta + \rho)} \right
) \right ] \; . ~~~~
%       (50)
\end{eqnarray}
Note that the term proportional to $m^{2}_{1} / \Delta m^{2}_{21}$
in the RGE of $\theta^{}_{12}$ in Eq. (47) is suppressed by
$\cos(\rho -\sigma) \simeq 0$ in this special case, and thus it has
been omitted from Eq. (50). The three mixing angles may therefore
receive comparably small radiative corrections for a modest value of
$\tan\beta$ (e.g., in the MSSM10 case). This observation was not
noticed in the literature simply because $\theta^{}_{13}$ used to be
assumed to be very small \cite{RGE11,RGE12,RGE13,RGE17}. If
$\tan\beta$ is sufficiently large (e.g., in the MSSM50 case),
however, the phase difference $(\rho - \sigma)$ will be able to
quickly run away from its initial value $(\rho -\sigma) \sim \pm
90^\circ$ due to the significant radiative corrections, implying
that Eq. (50) is no more a good approximation of Eq. (47).

Here let us consider a typical example of this special case --- the
correlative neutrino mixing pattern with $\theta^{}_{12} \simeq
35.3^\circ$, $\theta^{}_{23} = 45^\circ$ and $\theta^{}_{13} \simeq
9.7^\circ$ \cite{correlative}. Compared with the best-fit values of
the three mixing angles at $\Lambda^{}_{\rm EW}$, the three mixing
angles in this correlative mixing pattern at $\Lambda^{}_{\rm FS}$
have to receive comparably small radiative corrections during their
RGE evolution. As we have mentioned in the last paragraph, this
requirement can easily be achieved in the MSSM10 case provided the
condition $(\rho -\sigma) \simeq \pm 90^\circ$ is satisfied for a
nearly degenerate neutrino mass spectrum. Such a condition is
unnecessary if the neutrino mass spectrum has a strong hierarchy.
The numerical results are presented in Table 3, where $\delta =
-68^\circ$, $\rho = 13^\circ$ and $\sigma = 115^\circ$ are input at
$\Lambda_{\rm FS}$. Then
$\theta^{}_{12} = 34.52^\circ$, $\theta^{}_{23} = 45.98^\circ$ and
$\theta^{}_{13} = 8.83^\circ$ are obtained at $\Lambda_{\rm EW}$
after the radiative corrections.
%%%%%%%%%%%%%%%%%%%%%%%%%%%%%%%% Table 3 %%%%%%%%%%%%%%%%%%%%%%%%%
\begin{table}
\tbl{Radiative corrections to the correlative neutrino mixing
pattern from $\Lambda^{}_{\rm FS} \sim 10^{14}$ GeV to
$\Lambda^{}_{\rm EW} \sim 10^2$ GeV in the MSSM10.}
{\begin{tabular}{llllllll} ~ Parameter & ~~~~~~ & Input
($\Lambda^{}_{\rm FS}$) ~~~ & Output
($\Lambda^{}_{\rm EW}$) \\
\hline
~ $m^{}_{1} ~ ({\rm eV} )$ && 0.227 & 0.200 \\
~ $\Delta m^{2}_{21} ~ ( 10^{-5} ~{\rm eV}^2 )$ && 15.72 & 7.59 \\
~ $\Delta m^{2}_{31} ~ ( 10^{-3} ~{\rm eV}^2 )$ && 3.19 & 2.40 \\
~ $\theta^{}_{12}$ && $35.3^\circ$ & $34.52^\circ$ \\
~ $\theta^{}_{23}$ && $45^\circ$ & $45.98^\circ$ \\
~ $\theta^{}_{13}$ && $9.7^\circ$ & $8.83^\circ$ \\
~ $\delta$ && $-68^\circ$ & $-80.88^\circ$ \\
~ $\rho$ && $13^\circ$ & $19.64^\circ$ \\
~ $\sigma$ && $115^\circ$ & $118.03^\circ$ \\
~ $\delta + \rho + \sigma$ && $60^\circ$ & $56.79^\circ$ \\ \hline
\end{tabular}}
\end{table}
%%%%%%%%%%%%%%%%%%%%%%%%%%%%%%%%%%%%%%%%%%%%%%%%%%%%%%%%%%%%%%%%

(2) {\it The radiative generation of the CP-violating phases}.
It is well known that one CP-violating phase can be generated from
another \cite{LMX05}, simply because they are entangled in the RGEs.
An especially interesting example is the Dirac phase $\delta$, which
measures the strength of CP violation in neutrino oscillations at
the electroweak scale, can be radiatively generated from the nonzero
Majorana phases $\rho$ and $\sigma$ at a superhigh-energy scale. If
$\theta^{}_{13}$ is very small, however, the running of $\delta$ can
be significantly enhanced by the terms that are inversely
proportional to $\sin\theta^{}_{13}$. In the MSSM10 case it has been
found that even $\delta = 90^\circ$ can be radiatively generated if
$\theta^{}_{13} \simeq 1^\circ$ is taken \cite{LMX05}. Given
$\theta^{}_{13} \simeq 9^\circ$ at $\Lambda_{\rm
EW}$, it is found that $-30^\circ \leq \delta \leq 30^\circ$ at
$\Lambda_{\rm EW}$ may result from $\delta = 0^\circ$ at
$\Lambda^{}_{\rm FS}$ in the MSSM10 case. In the MSSM50 case even
$|\delta| \simeq 90^\circ$ can be obtained at $\Lambda_{\rm EW}$
\cite{LX12}.

(3) {\it The running of the sum $\delta + \rho + \sigma$}.
Eq. (48) leads us to the RGE of the sum of the three
CP-violating phases:
\begin{eqnarray}
\frac{{\rm d}}{{\rm d} t} (\delta + \rho + \sigma) & \approx &
\frac{y^{2}_{\tau}}{4 \pi^2} \frac{m^{2}_{1}}{\Delta m^{2}_{32}}
\frac{1}{s^{}_{13}} \left [ \frac{1}{2} c^{}_{12} s^{}_{12} c^{}_{23}
s^{}_{23}
\left ( s^{}_{(\delta + 2 \rho)} - s^{}_{(\delta + 2 \sigma)} \right )
\right. \nonumber\\[2mm]
& & ~~~  \left. - \left ( c^{}_{(\delta - \rho)} s^{}_{(\delta - \rho)}
s^{2}_{12} +c^{}_{(\delta - \sigma)} s^{}_{(\delta - \sigma)} c^{2}_{12}
\right ) \cos2\theta^{}_{23} s^{}_{13} \right. \nonumber\\[2mm]
& & ~~~ \left. + \left ( c^{}_{\rho} s^{}_{\rho} c^{2}_{12} +
c^{}_{\sigma} s^{}_{\sigma} s^{2}_{12} \right ) c^{2}_{23} s^{}_{13}
+ {\cal O} (s^{2}_{13}) \right ] \; .
%       (51)
\end{eqnarray}
Since the value of $\theta^{}_{13}$ is not small, the RGE running
effect on $(\delta + \rho + \sigma)$ is expected to be
insignificant. In other words, the sum of the three CP-violating
phases may approximately keep unchanged during the RGE evolution in the
SM or MSSM with a modest $\tan\beta$. The numerical
analysis shows that $(\delta + \rho + \sigma)$ changes less than
$4^\circ$ when running from $\Lambda^{}_{\rm FS}$ down to
$\Lambda^{}_{\rm EW}$ in the MSSM10 case. The stability of $(\delta
+ \rho + \sigma)$ against the radiative corrections is quite
impressive, unless $\tan\beta$ is sufficiently large \cite{LX12}.

To summarize, in order to obtain a phenomenologically-favored
neutrino mixing pattern at the electroweak scale $\Lambda^{}_{\rm
EW}$, the radiative corrections should be carefully examined for
those mixing patterns at a superhigh-energy scale $\Lambda^{}_{\rm
FS}$ which might result from a certain flavor symmetry. In the
MSSM10 case the values of $\theta^{}_{23}$ and $\theta^{}_{13}$
predicted at $\Lambda^{}_{\rm FS}$ are always close to their running
values at $\Lambda^{}_{\rm EW}$, while the value of $\theta^{}_{12}$
at $\Lambda^{}_{\rm FS}$ can be somewhat smaller or larger than its
running value at $\Lambda^{}_{\rm EW}$. In the MSSM50 case the
allowed ranges of the three mixing angles at $\Lambda^{}_{\rm FS}$
can be quite wide, as we have discussed above. However, a crucial
point is that a given flavor symmetry model should be able to
predict the appropriate CP-violating phases at $\Lambda^{}_{\rm FS}$
in order to obtain the appropriate mixing angles at $\Lambda^{}_{\rm
EW}$ after the RGE evolution. In general, it is possible to
generate $\theta^{}_{13} \simeq 9^\circ$ at $\Lambda^{}_{\rm EW}$
from $\theta^{}_{13} \simeq 0^\circ$ at $\Lambda^{}_{\rm FS}$
through the radiative corrections, in particular when some new degrees of
freedom or nontrivial running effects (such as the seesaw threshold
effects \cite{RGE04,RGE05,RGE17}) are taken into account or the three
CP-violating phases are fine-turned. Therefore, we argue that it seems
more natural for a specific flavor symmetry model to predict a
relatively large $\theta^{}_{13}$ at $\Lambda_{\rm FS}$.

A measurement of the Dirac phase $\delta$ in the forthcoming long-baseline
neutrino oscillation experiments and any experimental information about the
Majorana CP-violating phases $\rho$ and $\sigma$ are extremely
important, so as to distinguish one flavor symmetry model from
another through their different sensitivities to the radiative
corrections. This observation makes sense in particular after the
experimental errors associated with the neutrino mixing parameters
are comparable with or smaller than the magnitudes of their
respective RGE running effects.

\section{SEESAW-ENHANCED NEUTRINO DIPOLE MOMENTS}

The most popular mechanism of generating finite but tiny neutrino
masses beyond the SM is the canonical seesaw mechanism
\cite{SS1,SS2,SS3,SS4,SS5}, where the small neutrino masses are
attributed to the existence of heavy degrees of freedom such as the
right-handed Majorana neutrinos. In this elegant picture the
$3\times 3$ MNSP matrix $U$ has a striking
difference from the $3\times 3$ CKM matrix $V$ in the
SM: it is not exactly unitary due to small mixing between light and
heavy neutrinos as a result of the Yukawa interactions. The heavy
neutrinos can be searched for at the LHC if the
seesaw mechanism works at the TeV scale \cite{Xing09,LR1,LR2,LR3},
and the unitarity of $U$ can be tested in the future neutrino
oscillation experiments. Another
possibility is to look at the seesaw-induced non-unitary
effects on the electromagnetic dipole moments (EMDMs) and radiative
decays $\nu^{}_i \to \nu^{}_j + \gamma$
\cite{XZ12}. If the active neutrinos acquire their respective
masses in the seesaw mechanism, they should have the EMDMs through
quantum loops. The fact that the Majorana neutrinos are their own
antiparticles implies that they can only have the {\it transition}
EMDMs between two different neutrino mass eigenstates in an electric
or magnetic field. The relevant radiative decays of the heavier
active neutrinos, which may contribute to the cosmic infrared
background in the Universe \cite{CIB1,CIB2,CIB3}, are of particular
interest in cosmology.

\subsection{Analytical Discussions}

We focus on the seesaw-induced non-unitary effects on the EMDMs
and radiative decays of active Majorana neutrinos. The canonical
seesaw mechanism is based on a simple extension of the SM in which
three heavy right-handed neutrinos are added and the lepton number
is violated by their Majorana mass term \cite{SS1,SS2,SS3,SS4,SS5}:
\begin{eqnarray}
-{\cal L}_\nu = \overline{\ell^{}_{\rm L}} Y^{}_\nu \tilde{H}
N^{}_{\rm R} + \frac{1}{2} \overline{N^c_{\rm R}} M^{}_{\rm R}
N^{}_{\rm R} + {\rm h.c.} \; ,
%     (52)
\end{eqnarray}
where $\tilde{H} \equiv i \sigma^{}_2 H^*$ with $H$ being the SM
Higgs doublet, $\ell^{}_{\rm L}$ denotes the left-handed lepton
doublet, $N^{}_{\rm R}$ stands for the column vector of three
right-handed neutrinos, and $M^{}_{\rm R}$ is a symmetric Majorana
mass matrix. After spontaneous $SU(2)^{}_{\rm L} \otimes
U(1)^{}_{\rm Y} \to U(1)^{}_{\rm em}$ gauge symmetry breaking, $H$
achieves its vacuum expectation value $\langle H\rangle =
v/\sqrt{2}$ with $v \simeq 246$ GeV. Then the Yukawa-interaction
term in ${\cal L}^{}_\nu$ yields the Dirac mass matrix $M^{}_{\rm D}
= Y^{}_\nu v/\sqrt{2}$, but the Majorana mass term in ${\cal
L}^{}_\nu$ keeps unchanged since right-handed neutrinos are the
$SU(2)^{}_{\rm L}$ singlet and thus they are not subject to the
electroweak symmetry breaking. The overall neutrino mass matrix
turns out to be a symmetric $6\times 6$ matrix and can be
diagonalized through
\begin{eqnarray}
{\cal U}^\dagger \left( \begin{matrix} {\bf 0} & M^{}_{\rm D} \cr
M^T_{\rm D} & M^{}_{\rm R} \end{matrix} \right) {\cal U}^* = \left(
\begin{matrix} \widehat{M}^{}_\nu & {\bf 0} \cr {\bf 0} &
\widehat{M}^{}_N \end{matrix} \right) \; ,
%     (53)
\end{eqnarray}
where we have defined $\widehat{M}^{}_\nu \equiv {\rm
Diag}\{m^{}_1, m^{}_2, m^{}_3 \}$ and $\widehat{M}^{}_N \equiv {\rm
Diag}\{M^{}_1, M^{}_2, M^{}_3 \}$ with $m^{}_i$ and $M^{}_i$ being
the physical masses of three light neutrinos $\nu^{}_i$ and three
heavy neutrinos $N^{}_i$ (for $i=1,2,3$). The $6\times 6$ unitary
matrix $\cal U$ is decomposed as \cite{Xing3+3}
\begin{eqnarray}
{\cal U} = \left( \begin{matrix} {\bf 1} & {\bf 0} \cr {\bf 0} &
Z \cr \end{matrix} \right) \left( \begin{matrix} A & R \cr S &
B \cr \end{matrix} \right) \left(
\begin{matrix} X & {\bf 0} \cr {\bf 0} & {\bf 1} \cr
\end{matrix} \right) \; ,
%     (54)
\end{eqnarray}
where ${\bf 1}$ denotes the $3\times 3$ identity matrix, $X$
and $Z$ are the $3\times 3$ unitary matrices, and $A$, $B$, $R$
and $S$ are the $3\times 3$ matrices which characterize the
correlation between the active or light neutrino sector ($X$)
and the sterile or heavy neutrino sector ($Z$). A full
parametrization of $\cal U$ in terms of 15 mixing angles and 15
CP-violating phases has been given before \cite{Xing3+3}. One may
express the flavor eigenstates of three active neutrinos in terms of
the mass eigenstates $\nu^{}_i$ and $N^{}_i$. In the mass basis of
charged leptons and neutrinos, the leptonic weak charged-current
(cc) and neutral-current (nc) interactions read
\begin{eqnarray}
-{\cal L}^{}_{\rm cc} & = & \frac{g}{\sqrt{2}} \left[
\overline{l^{}_{\alpha \rm L}} \gamma^\mu \left( U^{}_{\alpha i}
\nu^{}_{i \rm L} + R^{}_{\alpha i} N^{}_{i
\rm L} \right) W^{-}_\mu + {\rm h.c.} \right] \; , \nonumber \\
-{\cal L}^{}_{\rm nc} & = & \frac{g}{2\cos\theta^{}_{\rm w}} \left\{
\overline{\nu^{}_{i \rm L}} \gamma^\mu (U^\dagger U)^{}_{ij}
\nu^{}_{j \rm L} + \overline{N^{}_{i \rm L}} \gamma^\mu (R^\dagger
R)^{}_{ij} N^{}_{j \rm L} \right. \nonumber\\
& & ~~~~~~ \left. + \left[\overline{\nu^{}_{i \rm L}}
\gamma^\mu (U^\dagger R)^{}_{ij} N^{}_{j \rm L} + {\rm h.c.} \right]
\right\} Z^{}_\mu \; ,
%     (55)
\end{eqnarray}
where $\alpha$ runs over $e$, $\mu$ or $\tau$, $U = AX$ is
responsible for the flavor mixing of active neutrinos $\nu^{}_i$,
and $R$ measures the strength of charged-current interactions of
heavy neutrinos $N^{}_i$ (for $i=1,2,3$) \cite{Xing2008}. A small
deviation of $U$ from $X$ is actually characterized by
nonvanishing $R$, as $UU^\dagger = AA^\dagger = {\bf 1} -
RR^\dagger$ holds. The exact seesaw relation between the masses of
light and heavy neutrinos is $U \widehat{M}^{}_\nu U^T + R
\widehat{M}^{}_N R^T = {\bf 0}$, which signifies the correlation
between neutrino masses and flavor mixing parameters.
%%%%%%%%%%%%%%%%%%%%%%%% Fig. 4 %%%%%%%%%%%%%%%%%%%%%%%%
\begin{figure}
  \begin{center}
  \includegraphics[width=0.8\textwidth]{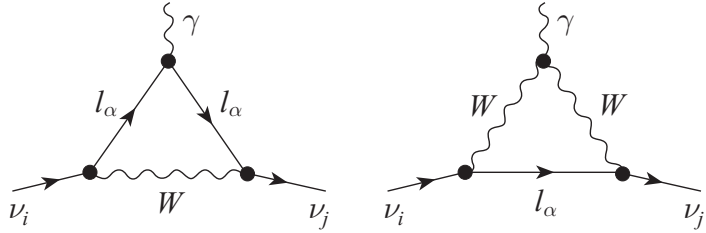} \\
  \end{center}
   \vspace{-0.3cm}
  \caption{The one-loop Feynman diagrams (and their charge-conjugate
  counterparts) contributing to the EMDMs of the Majorana neutrinos,
  where $\alpha = e, \mu, \tau$ and $i,j = 1,2,3$.}
\end{figure}
%%%%%%%%%%%%%%%%%%%%%%%%%%%%%%%%%%%%%%%%%%%%%%%%
%%%%%%%%%%%%%%%%%%%%%%%% Fig. 5 %%%%%%%%%%%%%%%%%%%%%%%%
\begin{figure}[t]
  \begin{center}
  \includegraphics[width=0.8\textwidth]{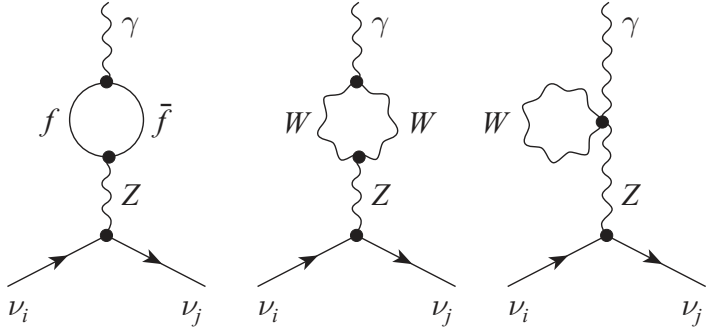} \\
  \end{center}
   \vspace{-0.3cm}
  \caption{The one-loop $\gamma$-$Z$ self-energy diagrams (and their
  charge-conjugate counterparts) associated with the EMDMs of massive Majorana
  neutrinos, where $f$ denotes all the SM fermions and $i,j = 1,2,3$.}
\end{figure}
%%%%%%%%%%%%%%%%%%%%%%%%%%%%%%%%%%%%%%%%%%%%%%%%

Let us consider the radiative $\nu^{}_i \to \nu^{}_j + \gamma$
transition, whose electromagnetic vertex can be written as
\begin{eqnarray}
\Gamma^\mu_{ij}(0) = \mu^{}_{ij} \left(i \ \sigma^{\mu\nu}
q^{}_\nu\right) + \epsilon^{}_{ij} \left(\sigma^{\mu\nu} q^{}_\nu
\gamma^{}_5 \right) \;
%     (56)
\end{eqnarray}
for a real photon satisfying the on-shell conditions $q^2 =0$ and
$q^{}_\mu \varepsilon^\mu =0$. In Eq. (56) $\epsilon^{}_{ij}$ and
$\mu^{}_{ij}$ are the electric and magnetic {\it transition} dipole
moments of Majorana neutrinos, and their sizes can be calculated via
the proper vertex diagrams in Fig. 4 (weak cc interactions). The
$\gamma$-$Z$ self-energy diagrams in Fig. 5 (weak nc interactions)
do not have any {\it net} contribution to $\epsilon^{}_{ij}$ and
$\mu^{}_{ij}$, but we find that they play a very crucial role in
eliminating the infinities because the divergent terms originating
from Fig. 4 are unable to automatically cancel out in the presence
of the seesaw-induced non-unitary effects (i.e., $R \neq {\bf 0}$
and $U \neq X$) unless those divergent terms originating from
Fig. 5 are also taken into account. This observation is new. It
implies that the non-unitary case under discussion is somewhat
different from the unitary case (i.e., $R={\bf 0}$ and $U^\dagger U
= X^\dagger X ={\bf 1}$) discussed before in the literature
\cite{Shrock,Wolfenstein,Kayser,Nieves}, where the Feynman diagrams
in Fig. 5 are forbidden and the divergent terms arising from Fig. 4
can automatically cancel out.

After a careful calculation, we arrive at
\begin{eqnarray}
\mu^{}_{ij} & = & \frac{i e G^{}_{\rm F}}{4\sqrt{2} \pi^2}
\left(m^{}_i + m^{}_j \right) \sum^{}_{\alpha} F^{}_\alpha \text{Im}
\left(U^{}_{\alpha i} U^*_{\alpha j} \right) \; ,
\nonumber \\
\epsilon^{}_{ij} & = & \frac{e G^{}_{\rm F}}{4\sqrt{2} \pi^2}
\left(m^{}_i - m^{}_j \right) \sum^{}_{\alpha} F^{}_\alpha \text{Re}
\left(U^{}_{\alpha i} U^*_{\alpha j} \right) \; ,
%     (57)
\end{eqnarray}
where
\begin{eqnarray}
F^{}_\alpha = \frac{3}{4} \left[
\frac{2-\xi^{}_\alpha}{1-\xi^{}_\alpha} - \frac{2
\xi^{}_\alpha}{\left(1-\xi^{}_\alpha\right)^2} + \frac{2
\xi^2_\alpha\ln \xi^{}_\alpha}{\left(1-\xi^{}_\alpha\right)^3}
\right] \;
%     (58)
\end{eqnarray}
with $\xi^{}_\alpha \equiv m^2_\alpha/M^2_W$ (for $\alpha = e, \mu,
\tau$) denotes the one-loop function. Although this result is {\it
formally} the same as that obtained in the literature
\cite{Shrock,Wolfenstein,Kayser,Nieves}, they are {\it
intrinsically} different as the seesaw-induced non-unitary effects
on $\mu^{}_{ij}$ and $\epsilon^{}_{ij}$ were not considered in the
previous works. To see how important such non-unitary effects may
be, let us make two analytical approximations. First, $F^{}_\alpha
\simeq 3 \left(2- \xi^{}_\alpha \right)/4$ holds to a good degree of
accuracy for $\xi^{}_\alpha \ll 1$. Second, $U = AX \simeq
X - T X$ is also a good approximation for small
non-unitary corrections to $X$, where \cite{Xing3+3}
\begin{eqnarray}
X & = & \left( \begin{matrix} c^{}_{12} c^{}_{13} & \hat{s}^*_{12}
c^{}_{13} & \hat{s}^*_{13} \cr -\hat{s}^{}_{12} c^{}_{23} -
c^{}_{12} \hat{s}^{}_{13} \hat{s}^*_{23} & c^{}_{12} c^{}_{23} -
\hat{s}^*_{12} \hat{s}^{}_{13} \hat{s}^*_{23} & c^{}_{13}
\hat{s}^*_{23} \cr \hat{s}^{}_{12} \hat{s}^{}_{23} - c^{}_{12}
\hat{s}^{}_{13} c^{}_{23} & -c^{}_{12} \hat{s}^{}_{23} -
\hat{s}^*_{12} \hat{s}^{}_{13} c^{}_{23} & c^{}_{13} c^{}_{23} \cr \end{matrix}
\right) ,
\nonumber \\
T & = & \left( \begin{matrix} \displaystyle\frac{1}{2} \sum^6_{k=4}
s^2_{1k} & 0 & 0 \cr \displaystyle\sum^6_{k=4} \hat{s}^{}_{1k}
\hat{s}^*_{2k} & \displaystyle\frac{1}{2} \sum^6_{k=4} s^2_{2k} & 0
\cr \displaystyle\sum^6_{k=4} \hat{s}^{}_{1k} \hat{s}^*_{3k} &
\displaystyle\sum^6_{k=4} \hat{s}^{}_{2k} \hat{s}^*_{3k} &
\displaystyle\frac{1}{2} \sum^6_{k=4} s^2_{3k} \cr \end{matrix} \right) ~
%     (59)
\end{eqnarray}
with $c^{}_{ij} \equiv \cos\theta^{}_{ij}$ and $\hat{s}^{}_{ij}
\equiv e^{i\delta^{}_{ij}} \sin\theta^{}_{ij}$ (here
$\theta^{}_{ij}$ and $\delta^{}_{ij}$ are the mixing angles and
CP-violating phases). Note that the light-heavy neutrino mixing
angles $\theta^{}_{ik}$ (for $i=1,2,3$ and $k=4,5,6$) are at most of
${\cal O}(0.1)$ \cite{Antusch}, such that the deviation of $U$ from
$X$ is at the percent level or much smaller. Then we obtain
\begin{eqnarray}
\sum_\alpha F^{}_\alpha \left(U^{}_{\alpha i} U^*_{\alpha j} \right)
& \simeq & -\frac{3}{2} \sum_\alpha \left[\left(X\right)^{}_{\alpha
i} \left(T X\right)^*_{\alpha j} + \left(T
X\right)^{}_{\alpha i} \left(X\right)^*_{\alpha j} \right] \nonumber\\
& & - \frac{3}{4} \sum_\alpha \left[ \xi^{}_\alpha
\left(X\right)^{}_{\alpha i} \left(X\right)^*_{\alpha j}
\right] \; .
%     (60)
\end{eqnarray}
The first and second terms on the right-hand side of this equation
correspond to the non-unitary and unitary contributions,
respectively. While the former is suppressed by $s^2_{ik} \lesssim
{\cal O}(10^{-2})$ (for $i=1,2,3$ and $k=4,5,6$) hidden in $T$, the
latter is suppressed by $\xi^{}_\alpha \lesssim 4.9 \times 10^{-4}$
(for $\alpha =e, \mu, \tau$) due to the GIM mechanism \cite{GIM}.
We therefore draw a generic conclusion that the seesaw-induced non-unitary
effects on $\epsilon^{}_{ij}$ and $\mu^{}_{ij}$ can be comparable
with or even larger than the standard (unitary) contributions.

In this case the rates of radiative $\nu^{}_i \to \nu^{}_j + \gamma$
decays are given by
\begin{eqnarray}
\Gamma^{}_{\nu^{}_i \to \nu^{}_j + \gamma} & = & \frac{\left(
m^2_i-m^2_j \right)^3}{8\pi m^3_i} \left(|\mu^{}_{ij}|^2_{} +
|\epsilon^{}_{ij}|^2_{} \right) \nonumber\\
& \simeq & 5.3 \times \left(1 -
\frac{m^2_j}{m^2_i} \right)^3 \left(\frac{m^{}_i}{{\rm 1 ~
eV}}\right)^3 \left(\frac{\mu^{}_{\rm eff}}{\mu^{}_{\rm B}}\right)^2
{\rm s}^{-1}
%     (61)
\end{eqnarray}
with $\mu^{}_{\rm eff} \equiv \sqrt{|\mu^{}_{ij}|^2 +
|\epsilon^{}_{ij}|^2}$ for $\nu^{}_i\to \nu^{}_j + \gamma$ being the
effective EMDMs and $\mu^{}_{\rm B} = e/(2m^{}_e)$ being the Bohr
magneton. The size of $\Gamma^{}_{\nu^{}_i \to \nu^{}_j + \gamma}$
can be experimentally constrained by observing no emission of the
photons from solar $\nu^{}_e$ and reactor $\overline{\nu}^{}_e$
fluxes. More stringent constraints on $\mu^{}_{\rm eff}$ come from
the Supernova 1987A limit on neutrino decays and from the
cosmological limit on distortions of the cosmic microwave background
radiation (in particular, its infrared part): $\mu^{}_{\rm eff} <
{\rm a ~ few} \times 10^{-11} ~\mu^{}_{\rm B}$
\cite{Raffelt99,Giunti09}. Now that more and more interest is being
paid to the cosmic infrared background relevant to the radiative
decays of massive neutrinos \cite{CIB1,CIB2,CIB3}, it is desirable
to evaluate $\mu^{}_{\rm eff}$ and $\Gamma^{}_{\nu^{}_i \to \nu^{}_j
+ \gamma}$ on a well-defined theoretical ground, such as the
canonical seesaw mechanism under discussion.
%%%%%%%%%%%%%%%%%%%%%%%% Fig. 6 %%%%%%%%%%%%%%%%%%%%%%%%
\begin{figure}
  \begin{center}
  \includegraphics[width=0.9\textwidth]{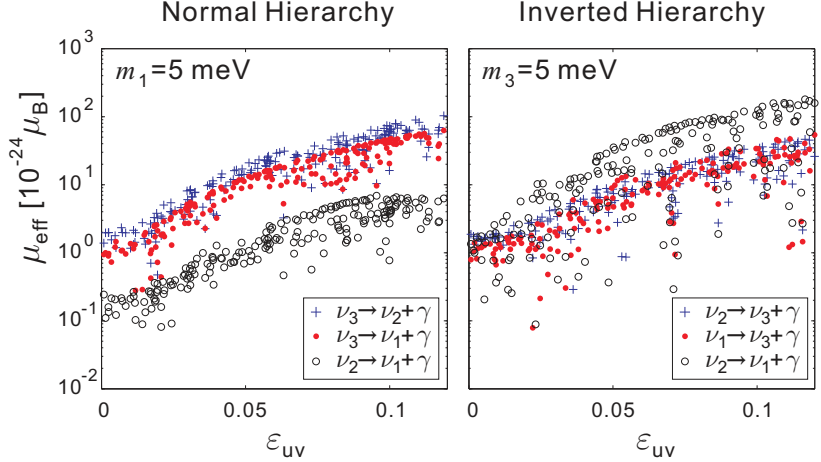} \\
  \end{center}
  \vspace{-0.3cm}
  \caption{Illustration of the seesaw-induced
  non-unitary effects on $\mu^{}_{\rm eff}$ for three active neutrinos.
  The standard (unitary) results correspond to $\varepsilon^{}_{\rm uv} =0$,
  and their uncertainties come from the three unknown CP-violating phases
  of $X$.}
\end{figure}
%%%%%%%%%%%%%%%%%%%%%%%%%%%%%%%%%%%%%%%%%%%%%%%%
%%%%%%%%%%%%%%%%%%%%%%%% Fig. 7 %%%%%%%%%%%%%%%%%%%%%%%%
\begin{figure}
  \begin{center}
  \includegraphics[width=0.9\textwidth]{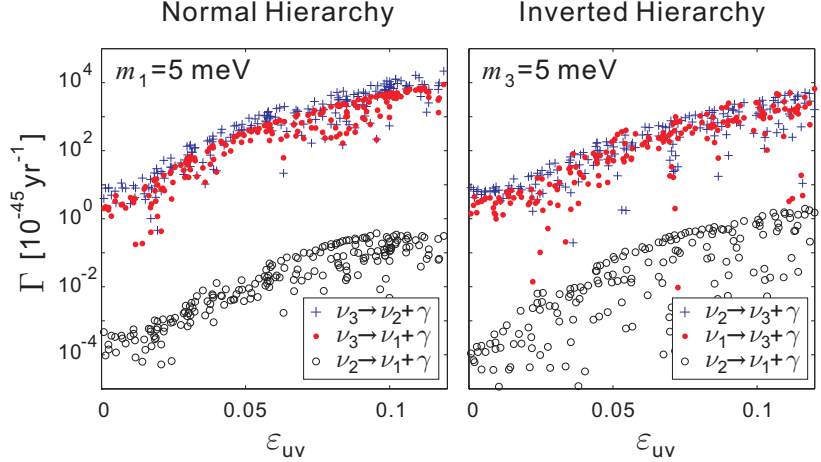} \\
  \end{center}
   \vspace{-0.3cm}
  \caption{Illustration of the seesaw-induced non-unitary
  effects on $\Gamma^{}_{\nu^{}_i \to \nu^{}_j + \gamma}$ for three active
  neutrinos. The standard (unitary) results correspond to
  $\varepsilon^{}_{\rm uv} =0$, and their uncertainties come from the three
  unknown CP-violating phases of $X$.}
\end{figure}
%%%%%%%%%%%%%%%%%%%%%%%%%%%%%%%%%%%%%%%%%%%%%%%%

\subsection{Numerical Illustration}

Figs. 6 and 7 illustrate the numerical results of the non-unitary
effects on $\mu^{}_{\rm eff}$ and $\Gamma^{}_{\nu^{}_i\to \nu^{}_j +
\gamma}$ respectively \cite{XZ12}. Note that in this numerical
analysis those small active-sterile neutrino mixing angles in Eq.
(59) are constrained by present experimental data \cite{Antusch} as
follows:
\begin{eqnarray}
&& T^{}_{11} < 5.5 \times 10^{-3} \; , ~~~~ \left| T^{}_{21} \right|
< 7.0 \times 10^{-5} \; ,
\nonumber \\
&& T^{}_{22} < 5.0 \times 10^{-3} \; , ~~~~ \left| T^{}_{31} \right|
< 1.6 \times 10^{-2} \; ,
\nonumber \\
&& T^{}_{33} < 5.0 \times 10^{-3} \; , ~~~~ \left| T^{}_{32} \right|
< 1.0 \times 10^{-2} \; .
%     (62)
\end{eqnarray}
All $s^{}_{ik}$ in $T$ (for $i=1,2,3$ and $k=4,5,6$) are positive or
vanishing. The CP phases $\delta^{}_{ik}$ are all allowed
to vary from zero to $2\pi$, but they must satisfy the above
constraints together with the relations $UU^\dagger + RR^\dagger =
{\bf 1}$ and $U \widehat{M}^{}_\nu U^T + R \widehat{M}^{}_N R^T =
{\bf 0}$. To assure that radiative corrections to the masses of
three light neutrinos (via the one-loop self-energy diagrams
involving the heavy neutrinos) are sufficiently small (e.g., smaller
than $0.5$ meV) and stable, we simply assume that the masses of
three heavy neutrinos are nearly degenerate \cite{Pilaftsis,Zhou}
and not more than ${\cal O}(1)$ TeV. This assumption implies that
the results shown here are for a limited and safe parameter space of
the TeV seesaw mechanism, but it is instructive enough to reveal the
salient features of the non-unitary effects on the effective EMDMs
$\mu^{}_{\rm eff} (\nu^{}_i\to \nu^{}_j + \gamma)$ and the radiative
decay rates $\Gamma^{}_{\nu^{}_i\to \nu^{}_j + \gamma}$.

To present our numerical results in a convenient way, let us define
\begin{eqnarray}
\varepsilon^{}_{\rm uv} \equiv \left[\sum^6_{k=4} \left( s^2_{1k} +
s^2_{2k} + s^2_{3k} \right)\right]^{1/2} \; ,
%     (63)
\end{eqnarray}
which measures the overall strength of the unitarity violation of
$U$, and $\varepsilon^{}_{\rm uv} \in [0, 0.15)$ is reasonably taken
in our calculations. Namely, we allow each $s^{}_{ik}$ (for
$i=1,2,3$ and $k=4,5,6$) to vary in the range $0 \leq s^{}_{ik} <
0.15$. The numerical dependence of $\mu^{}_{\rm eff} (\nu^{}_i\to
\nu^{}_j + \gamma)$ and $\Gamma^{}_{\nu^{}_i \to \nu^{}_j + \gamma}$
on $\varepsilon^{}_{\rm uv}$ is shown in Figs. 6 and 7,
respectively. Some discussions are in order.

(1) Switching off the non-unitary effects (i.e.,
$\varepsilon^{}_{\rm uv} =0$), we obtain the effective
electromagnetic dipole moments
\begin{eqnarray}
\mu^{}_{\rm eff} \simeq \left\{ \begin{array}{l} \left(0.8 \sim
3.0\right) \times 10^{-25} ~\mu^{}_{\rm B} ~~~~~~ (\nu^{}_2 \to
\nu^{}_1 + \gamma) \; , \\
\left(0.8 \sim 1.5\right) \times 10^{-24} ~\mu^{}_{\rm B} ~~~~~~
(\nu^{}_3 \to \nu^{}_1 + \gamma) \; , \\
\left(1.1 \sim 2.1\right) \times 10^{-24} ~\mu^{}_{\rm B} ~~~~~~
(\nu^{}_3 \to \nu^{}_2 + \gamma) \; , \end{array} \right .
%     (64)
\end{eqnarray}
for the normal mass hierarchy with $m^{}_1 \simeq 5$ meV; and
\begin{eqnarray}
\mu^{}_{\rm eff} \simeq \left\{ \begin{array}{l} \left(0.01 \sim 2.0
\right) \times 10^{-24} ~\mu^{}_{\rm B} ~~~~~ (\nu^{}_2
\to \nu^{}_1 + \gamma) \; , \\
\left(0.8 \sim 1.5\right) \times 10^{-24} ~\mu^{}_{\rm B} ~~~~~~\;
(\nu^{}_3 \to \nu^{}_1
+ \gamma) \; , \\
\left(1.3 \sim 2.0\right) \times 10^{-24} ~\mu^{}_{\rm B} ~~~~~~\;
(\nu^{}_3 \to \nu^{}_2 + \gamma) \; , \end{array} \right .
%     (65)
\end{eqnarray}
for the inverted mass hierarchy with $m^{}_3 \simeq 5$ meV, where
the uncertainties mainly come from the unknown CP phases
$\delta^{}_{12}$, $\delta^{}_{13}$ and $\delta^{}_{23}$. Such
standard (unitary) results are far below the observational upper
bound on $\mu^{}_{\rm eff}$ ($<$ a few $\times 10^{-11} ~\mu^{}_{\rm
B}$ \cite{Raffelt99,Giunti09}), but they serve as a good reference
to the non-unitary effects on $\mu^{}_{\rm eff}$ being explored.

(2) Figs. 6 and 7 clearly show that $\mu^{}_{\rm eff}$ and
$\Gamma^{}_{\nu^{}_i \to \nu^{}_j + \gamma}$ can be maximally
enhanced by a factor of ${\cal O}(10^2)$ and a factor of ${\cal
O}(10^4)$, respectively, in particular when $\varepsilon^{}_{\rm
uv}$ approaches its upper limit as set by current experimental data.
The magnitude of $\mu^{}_{\rm eff} (\nu^{}_2 \to \nu^{}_1 + \gamma)$
may be strongly suppressed in the inverted neutrino mass hierarchy.
The reason is rather simple: on the one hand, $m^{}_1 \simeq m^{}_2$
holds in this case, and thus $\epsilon^{}_{12} \propto (m^{}_2 -
m^{}_1)$ must be very small; on the other hand, $\mu^{}_{12}$
depends on ${\rm Im}(U^{}_{\alpha 1} U^*_{\alpha 2})$, so it can
also be very small when the CP-violating phases are around zero or
$\pi$. This two-fold suppression becomes severer for the decay rate
$\Gamma^{}_{\nu^{}_2 \to \nu^{}_1 + \gamma}$, because it is
proportional to $ (m^{}_2 - m^{}_1)^3 \mu^{2}_{\rm eff} (\nu^{}_2
\to \nu^{}_1 + \gamma)$.

(3) The results of $\mu^{}_{\rm eff}$ and $\Gamma^{}_{\nu^{}_i \to
\nu^{}_j + \gamma}$ are sensitive to the absolute neutrino mass
scale for both normal and inverted mass hierarchies. For instance,
$\mu^{}_{\rm eff} (\nu^{}_2 \to \nu^{}_1 + \gamma)$ and $\mu^{}_{\rm
eff} (\nu^{}_3 \to \nu^{}_1 + \gamma)$ get enhanced when $m^{}_1$
changes from zero to 5 meV in the normal mass hierarchy; while
$\mu^{}_{\rm eff} (\nu^{}_1 \to \nu^{}_3 + \gamma)$ and $\mu^{}_{\rm
eff} (\nu^{}_2 \to \nu^{}_3 + \gamma)$ are enhanced when $m^{}_3$
changes from zero to 5 meV in the inverted mass hierarchy. This kind
of sensitivity is not so obvious if one only takes a look at the
expressions of $\mu^{}_{ij}$ and $\epsilon^{}_{ij}$ in Eq. (57). The
main reason is that a change of $m^{}_1$ or $m^{}_3$ requires some
fine-tuning of the active-sterile neutrino mixing angles and
CP-violating phases as dictated by the exact seesaw relation $U
\widehat{M}^{}_\nu U^T + R \widehat{M}^{}_N R^T = {\bf 0}$, leading
to a possibly significant change of $\mu^{}_{\rm eff}$. The
dependence of $\Gamma^{}_{\nu^{}_i \to \nu^{}_j + \gamma}$ on the
absolute neutrino mass scale is somewhat more complicated, as one
can see from Eq. (61).

(4) The CP-violating phases play a very important role in fitting
both the exact seesaw relation and Eq. (62). If the heavy neutrino
masses $M^{}_i$ are not suppressed, then an appreciable value of
$\varepsilon^{}_{\rm uv}$ requires some fine cancellations in the
matrix product $R \widehat{M}^{}_N R^T$ such that sufficiently small
$m^{}_i$ can be obtained from $U \widehat{M}^{}_\nu U^T = -R
\widehat{M}^{}_N R^T$. On the other hand, we remark that it is
actually unnecessary to require $M^{}_i$ to be around or above the
electroweak scale. The seesaw-induced non-unitary effects on
$\mu^{}_{\rm eff}$ and $\Gamma^{}_{\nu^{}_i \to \nu^{}_j + \gamma}$
can be significant even if one allows one, two or three heavy
neutrinos to be relatively light (e.g., at the keV mass scale). Such
sterile neutrinos are interesting in particle physics and cosmology.
Note that it is easier to satisfy the exact seesaw relation with an
appreciable value of $\varepsilon^{}_{\rm uv}$ by arranging $M^{}_i$
to lie in the keV, MeV or GeV range. This kind of low-scale seesaw
scenarios \cite{DeGouvea} might be technically natural, but they
have more or less lost the seesaw spirit. Of course, sufficiently
large $M^{}_i$ and sufficiently small $\theta^{}_{ik}$ can always
coexist to make the seesaw mechanism work in a natural way, but in
this traditional case the non-unitary effects are too small to have
any measurable consequences at low energies.

It is also worth pointing out that the seesaw-induced non-unitary
effects on $\mu^{}_{ij}$ and $\epsilon^{}_{ij}$ are rather different
from the case of making a naive assumption of the flavor mixing
between three active neutrinos and a few light sterile neutrinos
\cite{sterile review}. The latter can directly break the unitarity
of the $3\times 3$ MNSP matrix $U$ and then lift
the GIM suppression \cite{GIM} associated with $\mu^{}_{ij}$ and
$\epsilon^{}_{ij}$. This kind of non-unitary effects are not
constrained by the seesaw relation, and thus they are more arbitrary
and less motivated from the point of view of model building.

The effective electromagnetic dipole moments of three neutrinos and
the rates of their radiative decays can be maximally enhanced by a
factor of ${\cal O}(10^2)$ and a factor of ${\cal O}(10^4)$,
respectively, no matter whether the seesaw scale is around or below
the TeV energy scale. This observation is new and nontrivial, and it
reveals an intrinsic and presumably important correlation
between the electromagnetic properties of neutrinos and the
origin of their masses. Such a correlation may even serve as a
sensitive touch-stone for the highly-regarded seesaw mechanism.

\section{SUMMARY}

After the Daya Bay measurement of the smallest neutrino mixing angle
$\theta^{}_{13}$, it is natural to ask where we are standing and
where we are expecting to go in neutrino physics. We have tried to
answer these two questions from a phenomenological point of view in
this review paper, although our answers are incomplete and full of
conjectures. To be specific, we have given a fast overview of some
fundamental neutrino properties and paid particular interest to the
flavor issues of charged leptons and neutrinos, including the mass
spectrum, flavor mixing pattern and CP violation. We have gone into
details of possible lepton flavor structures by describing two
useful phenomenological strategies and giving a number of typical
examples. The impact of large $\theta^{}_{13}$ on the running
behaviors of other flavor mixing parameters has been discussed in
the framework of the MSSM. We have also illustrated the
seesaw-enhanced electromagnetic dipole moments of three Majorana
neutrinos based on a viable TeV seesaw scenario.

%%%%%%%%%%%%%%%%%%%%%%%% Fig. 8 %%%%%%%%%%%%%%%%%%%%%%%%
\begin{figure}
\centering
\includegraphics[width=0.6\textwidth]{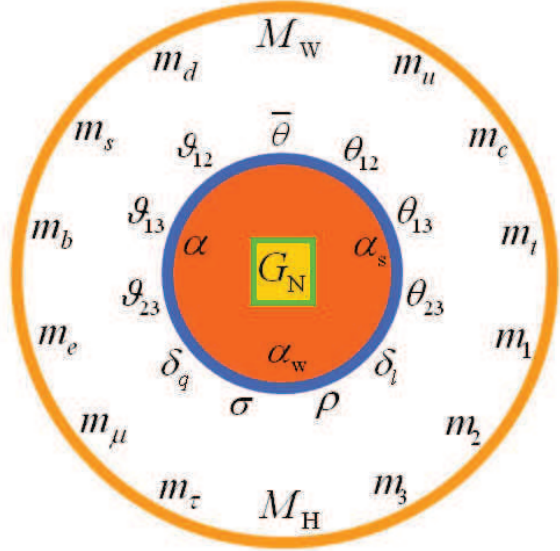}
 \vspace{0.3cm}
\caption{The Fritzsch-Xing plot for 29 fundamental parameters in
Nature, describing four kinds of interactions ($G^{}_{\rm N}$,
$\alpha$, $\alpha^{}_{\rm w}$ and $\alpha^{}_{\rm s}$), six quark
masses ($m^{}_u$, $m^{}_c$, $m^{}_t$ and $m^{}_d$, $m^{}_s$,
$m^{}_b$), six lepton masses ($m^{}_e$, $m^{}_\mu$, $m^{}_\tau$ and
$m^{}_1$, $m^{}_2$, $m^{}_3$), three quark flavor mixing angles
($\vartheta^{}_{12}$, $\vartheta^{}_{13}$ and $\vartheta^{}_{23}$),
three lepton flavor mixing angles ($\theta^{}_{12}$,
$\theta^{}_{13}$ and $\theta^{}_{23}$), two CP-violating phases in
the quark sector ($\delta^{}_q$ and $\overline{\theta}$), three
CP-violating phases in the lepton sector ($\delta^{}_l$, $\rho$ and
$\sigma$), the Higgs mass $M^{}_{\rm H}$ and the $W$-boson mass
$M^{}_{\rm W}$.}
\end{figure}
%%%%%%%%%%%%%%%%%%%%%%%%%%%%%%%%%%%%%%%%%%%%%%%%

If only the SM particles are taken into account and the
massive neutrinos are assumed to be the Majorana particles, we are
then left with 29 fundamental parameters in Nature, as described by
the so-called Fritzsch-Xing plot in Fig. 8. The determination of
$\theta^{}_{13}$ and $M^{}_{\rm H}$ in 2012 is a milestone in
particle physics. But the effective strong CP-violating phase
$\overline{\delta}$ in the quark sector and the three weak
CP-violating phases (i.e., $\delta^{}_l$, $\rho$ and $\sigma$) in
the lepton sector remain unknown. Moreover, the absolute mass scale
of three neutrinos and their mass ordering have not been determined.
We hope that various precision neutrino experiments can help pin
down the relevant parameters in the lepton sector and then shed
light on the flavor dynamics in the foreseeable future.

Of course, the picture in Fig. 8 may be too simple and too naive
because we are not sure whether some of those ``fundamental"
parameters are really fundamental or not. New degrees of freedom,
such as the sterile neutrinos or the supersymmetric particles, might
be discovered and make the flavor sector much messier. If the
underlying flavor theory is regarded as an animal, one has no idea
whether it is a donkey or an elephant or something else. In this
sense we are blind today and have to make a lot of experimental and
theoretical efforts to identify its nose, eyes, ears, legs and so on
in order to make sure what animal it is. The road behind has
repeatedly told us that the road ahead is always challenging, but it
is always exciting.

\vspace{0.5cm}

This review paper is essentially based on the plenary talk given by
one of us (Z.Z.X.) at the {\it SUSY 2012} conference. The work of
S.L. is supported in part by the National Basic Research Program
(973 Program) of China under Grant No. 2009CB824800, the National
Natural Science Foundation of China under Grant No. 11105113 and the
Fujian Provincial Natural Science Foundation under Grant No.
2011J05012. The work of Z.Z.X. is supported in part by the National
Natural Science Foundation of China under grant No. 11135009.


\begin{thebibliography}{99}

\bibitem{DYB} F.P. An {\it et al.} (Baya Bay Collaboration),
Phys. Rev. Lett. {\bf 108}, 171803
(2012).

\bibitem{T2K} K. Abe {\it et al.}  (T2K Collaboration),
Phys. Rev. Lett.  {\bf 107}, 041801 (2011).

\bibitem{MINOS} P. Adamson {\it et al.}  (MINOS Collaboration),
Phys. Rev. Lett. {\bf 107}, 181802 (2011).

\bibitem{Double Chooz} Y. Abe {\it et al.}  (Double Chooz Collaboration),
Phys. Rev. Lett. {\bf 108}, 131801 (2012).

\bibitem{ATLAS} G. Aad {\it et al.}  (ATLAS Collaboration),
Phys. Lett. B {\bf 716}, 1 (2012).

\bibitem{CMS} S. Chatrchyan {\it et al.}  (CMS Collaboration),
Phys. Lett. B {\bf 716}, 30 (2012).

\bibitem{Einstein} A. Einstein, Ann. Phys. {\bf 17}, 891 (1905).

\bibitem{tachyon} G. Feinberg, Phys. Rev. {\bf 159}, 1089 (1967).

\bibitem{OPERA} T. Adam {\it et al.}  (OPERA Collaboration), arXiv:1109.4897 [hep-ex].

\bibitem{Fermilab1} J. Alspector {\it et al.}, Phys. Rev. Lett. {\bf 36}, 837 (1976).

\bibitem{Fermilab2} G.R. Kalbfleisch  {\it et al.}, Phys. Rev. Lett. {\bf 43}, 1361 (1979).

\bibitem{MINOS07} P. Adamson {\it et al.}  (MINOS Collaboration),
Phys. Rev. D {\bf 76}, 072005 (2007).

\bibitem{ICARUS1} M. Antonello {\it et al.}  (ICARUS Collaboration),
Phys. Lett. B {\bf 713}, 17 (2012).

\bibitem{ICARUS2} M. Antonello {\it et al.}, arXiv:1208.2629 [hep-ex].

\bibitem{Borexino} P. Alvarez Sanchez {\it et al.}  (Borexino Collaboration),
Phys. Lett. B {\bf 716}, 401 (2012).

\bibitem{LVD} N.Y. Agafonova {\it et al.}  (LVD Collaboration),
Phys. Rev. Lett. {\bf 109}, 070801 (2012).

\bibitem{SN1987A1} K. Hirata {\it et al.}  (Kamiokande II Collaboration),
Phys. Rev. Lett. {\bf 58}, 1490 (1987).

\bibitem{SN1987A2} M.J. Longo, Phys. Rev. D {\bf 36}, 3276 (1987).

\bibitem{SN1987A3} L. Stodolsky, Phys. Lett. B {\bf 201}, 353 (1988).

\bibitem{SS1} P. Minkowski, Phys. Lett. B {\bf 67}, 421 (1977).

\bibitem{SS2} T. Yanagida, in {\it Proceedings of the Workshop on
Unified Theory and the Baryon Number of the Universe},
edited by O. Sawada and A. Sugamoto (KEK, Tsukuba, 1979), p. 95.

\bibitem{SS3} M. Gell-Mann, P. Ramond, and R. Slansky, in {\it Supergravity},
edited by P. van Nieuwenhuizen and D. Freedman (North Holland,
Amsterdam, 1979), p. 315.

\bibitem{SS4} S.L. Glashow, in {\it Quarks and Leptons},
edited by M. L$\acute{\rm e}$vy {\it et al.} (Plenum, New York, 1980), p. 707.

\bibitem{SS5} R.N. Mohapatra and G. Senjanovic,
Phys. Rev. Lett. {\bf 44}, 912 (1980).

\bibitem{keV1} S. Dodelson and L.M. Widrow,
Phys. Rev. Lett. {\bf 72}, 17 (1994).

\bibitem{keV2} X.D. Shi and G.M. Fuller, Phys. Rev. Lett. {\bf 82}, 2832 (1999)

\bibitem{keV3} A.D. Dolgov and S.H. Hansen, Astropart.
Phys. {\bf 16}, 339 (2002).

\bibitem{keV4} F. Bezrukov, H. Hettmansperger, and M. Lindner,
Phys. Rev. D {\bf 81}, 085032 (2010).

\bibitem{keV5} W. Liao, Phys. Rev. D {\bf 82}, 073001 (2010).

\bibitem{keV6} Y.F. Li and Z.Z. Xing, Phys. Lett. B {\bf 695}, 205 (2011).

\bibitem{keV7} M. Nemevsek, G. Senjanovic and Y. Zhang, JCAP {\bf 1207}, 006 (2012).

\bibitem{Xing03} Z.Z. Xing, Phys. Rev. D {\bf 68}, 053002 (2003).

\bibitem{Xing04} Z.Z. Xing, Int. J. Mod. Phys. A {\bf 19}, 1 (2004).

\bibitem{Rodejohann11} W. Rodejohann,
Int. J. Mod. Phys. E {\bf 20}, 1833 (2011).

\bibitem{Rodejohann12} W. Rodejohann, arXiv:1206.2560 [hep-ph].

\bibitem{Shrock77} B.W. Lee and R. Shrock, Phys. Rev. D {\bf 16}, 1444 (1977).

\bibitem{Shrock80} K. Fujikawa and R. Shrock, Phys. Rev. Lett. {\bf 45}, 963 (1980).

\bibitem{Vissani} A. Strumia and F. Vissani, hep-ph/0606054.

\bibitem{sterile review} K.N. Abazajian {\it et al.}, arXiv:1204.5379 [hep-ph].

\bibitem{FY} M. Fukugita and T. Yanagida, Phys. Lett. B {\bf 174}, 45 (1986).

\bibitem{LSND} A. Aguilar {\it et al.} (LSND Collaboration),
Phys. Rev. D {\bf 64}, 112007 (2001).

\bibitem{MiniBooNE} A.A. Aguilar-Arevalo {\it et al.} (MiniBooNE Collaboration),
Phys. Rev. Lett. {\bf 105}, 181801 (2010).

\bibitem{reactor} G. Mention {\it et al.}, Phys. Rev. D {\bf  83}, 073006 (2011).

\bibitem{Schwetz} J. Kopp, M. Maltoni, and T. Schwetz,
Phys. Rev. Lett. {\bf 107}, 091801 (2011).

\bibitem{Giunti} C. Giunti and M. Laveder, Phys. Rev. D {\bf 84}, 073008 (2011).

\bibitem{Mangano} See, e.g., G. Mangano and P.D. Serpico,
Phys. Lett. B {\bf 701}, 296 (2011); and references therein.

\bibitem{Raffelt1} J. Hamann {\it et al.}, Phys. Rev. Lett. {\bf 105}, 181301 (2010).

\bibitem{Raffelt2} J. Hamann {\it et al.}, JCAP {\bf 1109}, 034 (2011).

\bibitem{Raffelt3} E. Giusarma {\it et al.}, Phys. Rev. D {\bf 83}, 115023 (2011).

\bibitem{Bode} P. Bode, J.P. Ostriker, and N. Turok, Astrophys. J. {\bf 556}, 93 (2001).

\bibitem{Xing3+3} Z.Z. Xing, Phys. Rev. D {\bf 85}, 013008 (2012).

\bibitem{MNS1} Z. Maki, M. Nakagawa, and S. Sakata,
Prog. Theor. Phys. {\bf 28}, 870 (1962).

\bibitem{MNS2} B. Pontecorvo, Sov. Phys. JETP {\bf 26}, 984 (1968).

\bibitem{PDG} J. Beringer et al. (Particle Data Group),
Phys. Rev. D {\bf 86}, 010001 (2012).

\bibitem{Fogli} G.L. Fogli {\it et al.}, Phys. Rev. D {\bf 86}, 013012 (2012).

\bibitem{GG} M.C. Gonzalez-Garcia and M. Maltoni,
Phys. Rept. {\bf 460}, 1 (2008); and references therein.

\bibitem{hierarchy1} S. Choubey, S.T. Petcov, and M. Piai,
Phys. Rev. D {\bf 68}, 113006 (2003).

\bibitem{hierarchy2} L. Zhan, Y. Wang, J. Cao and L. Wen,
Phys. Rev. D {\bf 78}, 111103 (2008).

\bibitem{MSM1} P.H. Frampton, S.L. Glashow, and T. Yanagida,
Phys. Lett. B {\bf 548}, 119 (2002).

\bibitem{MSM2} For a review, see: W.L. Guo, Z.Z. Xing, and S. Zhou,
Int. J. Mod. Phys. E {\bf 16}, 1 (2007).

\bibitem{FL1} R. Friedberg and T.D. Lee,
High Energy Phys. Nucl. Phys. {\bf 30}, 591 (2006).

\bibitem{FL2} Z.Z. Xing, H. Zhang, and S. Zhou, Phys. Lett. B {\bf 641}, 189 (2006).

\bibitem{FL3} S. Luo and Z.Z. Xing, Phys. Lett. B {\bf 646}, 242 (2007).

\bibitem{FL4} Z.Z. Xing, Int. J. Mod. Phys. E {\bf 16}, 1361 (2007).

\bibitem{FL5} C. Jarlskog, Phys. Rev. D {\bf 77}, 073002 (2008).

\bibitem{FL6} C.S. Huang, T.J. Li, W. Liao, and S.H. Zhu,
Phys. Rev. D {\bf 78}, 013005 (2008).

\bibitem{FL7} S. Luo, Z.Z. Xing, and X. Li, Phys. Rev. D {\bf 78}, 117301 (2008).

\bibitem{XZ08} Z.Z. Xing and S. Zhou, Phys. Lett. B {\bf 666}, 166 (2008).

\bibitem{J} C. Jarlskog, Phys. Rev. Lett. {\bf 55}, 1039 (1985).

\bibitem{Xing96} Z.Z. Xing, Nuovo Cim. A {\bf 109}, 115 (1996)

\bibitem{Xing97} Z.Z. Xing, J. Phys. G {\bf 23}, 717 (1997).

\bibitem{NewX} Z.Z. Xing, arXiv:1210.1523 [hep-ph].

\bibitem{FX00} H. Fritzsch and Z.Z. Xing,
Prog. Part. Nucl. Phys. {\bf 45}, 1 (2000); and references therein.

\bibitem{Xing2012} Z.Z. Xing, Chin. Phys. C {\bf 36}, 281 (2012).

\bibitem{FX2000} H. Fritzsch and Z.Z. Xing, Phys. Rev. D {\bf 61}, 073016 (2000).

\bibitem{XZZ1} Z.Z. Xing, H. Zhang, and S. Zhou, Phys. Rev. D {\bf 77}, 113016 (2008).

\bibitem{XZZ2} Z.Z. Xing, H. Zhang, and S. Zhou, Phys. Rev. D {\bf 86}, 013013 (2012).

\bibitem{FX06} H. Fritzsch and Z.Z. Xing, Phys. Lett. B {\bf 634}, 514 (2006).

\bibitem{FX09} H. Fritzsch and Z.Z. Xing, Phys. Lett. B {\bf 682}, 220 (2009).

\bibitem{Fritzsch78} H. Fritzsch, Phys. Lett. B {\bf 73}, 317 (1978).

\bibitem{Fritzsch79} H. Fritzsch, Nucl. Phys. B {\bf 155}, 189 (1979).

\bibitem{Xing02} Z.Z. Xing, Phys. Lett. B {\bf 550}, 178 (2002).

\bibitem{ZX05} S. Zhou and Z.Z. Xing, Eur. Phys. J. C {\bf 38}, 495 (2005);

\bibitem{FXZ11} H. Fritzsch, Z.Z. Xing, Y.L. Zhou,
Phys. Lett. B {\bf 697}, 357 (2011).

\bibitem{Zero1} Z.Z. Xing, Phys. Lett. B {\bf 530}, 159 (2002).

\bibitem{Zero2} P.H. Frampton, S.L. Glashow, and D. Marfatia,
Phys. Lett. B {\bf 536}, 79 (2002).

\bibitem{Zero3} Z.Z. Xing, Phys. Lett. B {\bf 539}, 85 (2002).

\bibitem{Zero4} For a systematic analysis, see: H. Fritzsch, Z.Z. Xing, and S. Zhou,
JHEP {\bf 1109}, 083 (2011).

\bibitem{FN} C.D. Froggatt and H.B. Nielsen, Nucl. Phys. B {\bf 147}, 277 (1979).

\bibitem{Grimus} See, e.g., W. Grimus, A.S. Joshipura, L. Lavoura, and M. Tanimoto,
Eur. Phys. J. C {\bf 36}, 227 (2004).

\bibitem{FX96} H. Fritzsch and Z.Z. Xing, Phys. Lett. B {\bf 372}, 265 (1996).

\bibitem{FX98} H. Fritzsch and Z.Z. Xing, Phys. Lett. B {\bf 440}, 313 (1998).

\bibitem{Flavor1} G. Altarelli and F. Feruglio, Rev. Mod. Phys. {\bf 82}, 2701 (2010).

\bibitem{Flavor2} L. Merlo, arXiv:1004.2211 [hep-ph].

\bibitem{BM1} F. Vissani, hep-ph/9708483.

\bibitem{BM2} V. Barger, S. Pakvasa, T.J. Weiler, and K. Whisnant,
Phys. Lett. B {\bf 437}, 107 (1998).

\bibitem{TB1} P.F. Harrison, D.H. Perkins, and W.G. Scott,
Phys. Lett. B {\bf 530}, 167 (2002).

\bibitem{TB2} Z.Z. Xing, Phys. Lett. B {\bf 533}, 85 (2002).

\bibitem{TB3} P.F. Harrison and W.G. Scott, Phys. Lett. B {\bf 535}, 163 (2002).

\bibitem{TB4} X.G. He and A. Zee, Phys. Lett. B {\bf 560}, 87 (2003).

\bibitem{GR1} Y. Kajiyama, M. Raidal, and A. Strumia,
Phys. Rev. D {\bf 76}, 117301 (2007).

\bibitem{GR2} A slight variation of this golden-ratio mixing pattern
has been discussed by W. Rodejohann, Phys. Lett. B {\bf 671}, 267 (2009).

\bibitem{hexagonal1} Z.Z. Xing, J. Phys. G {\bf 29}, 2227 (2003).

\bibitem{hexagonal2} C. Giunti, Nucl. Phys. B (Proc. Suppl.) {\bf 117}, 24 (2003).

\bibitem{hexagonal3} The name of this flavor mixing pattern was coined
by C.H. Albright, A. Dueck, and W. Rodejohann, Eur. Phys. J. C {\bf 70}, 1099 (2010).

\bibitem{correlative} Z.Z. Xing, Phys. Lett. B {\bf 696}, 232 (2011).

\bibitem{RZZ} W. Rodejohann, H. Zhang, and S. Zhou,
Nucl. Phys. B {\bf 855}, 592 (2012).

\bibitem{Feruglio} R. de Adelhart Toorop, F. Feruglio,
and C. Hagedorn, Nucl. Phys. B {\bf 858}, 437 (2012).

\bibitem{Xing08} Z.Z. Xing, Phys. Rev. D {\bf 78}, 011301 (2008).

\bibitem{RGE01} S. Antusch, J. Kersten, M. Lindner, and
M. Ratz, Phys. Lett. B {\bf 544}, 1 (2002).

\bibitem{RGE02} S. Antusch and M. Ratz, JHEP {\bf 0211}, 010 (2002).

\bibitem{RGE03} J.W. Mei and Z.Z. Xing, Phys. Rev. D {\bf 70}, 053002 (2004).

\bibitem{RGE04} J.W. Mei and Z.Z. Xing, Phys. Lett. B {\bf 623}, 227 (2005).

\bibitem{RGE05} J.W. Mei, Phys. Rev. D {\bf 71}, 073012 (2005).

\bibitem{RGE06} S. Luo and Z.Z. Xing, Phys. Lett. B {\bf 632}, 341 (2006).

\bibitem{RGE07} S. Luo and Z.Z. Xing, Phys. Lett. B {\bf 637}, 279 (2006).

\bibitem{RGE08} S. Goswami, S.T. Petcov, S. Ray, and W. Rodejohann,
Phys. Rev. D {\bf 80}, 053013 (2009).

\bibitem{RGE09} T. Araki, C.Q. Geng, and Z.Z. Xing,
Phys. Lett. B {\bf 699}, 276 (2011).

\bibitem{RGE10} T. Araki and C.Q. Geng, JHEP {\bf 1109}, 139 (2011).

\bibitem{ZZ} H. Zhang and S. Zhou, Phys. Lett. B {\bf 704}, 296 (2011).

\bibitem{LX12}  S. Luo and Z.Z. Xing, arXiv:1203.3118 [hep-ph].

\bibitem{XZ07} Z.Z. Xing and S. Zhou, Phys. Lett. B {\bf 653}, 278 (2007).

\bibitem{Zhou12} S. Zhou, arXiv:1205.0761 [hep-ph].

\bibitem{Xing12} Z.Z. Xing, Chin. Phys. C {\bf 36}, 101 (2012).

\bibitem{RGE11} P.H. Chankowski and Z. Pluciennik, Phys. Lett.  B {\bf 316}, 312 (1993).

\bibitem{RGE12} K.S. Babu, C.N. Leung, and J.T. Pantaleone,
Phys. Lett. B {\bf 319}, 191 (1993).

\bibitem{RGE13} N. Haba, N. Okamura, and M. Sugiura,
Prog. Theor. Phys. {\bf 103}, 367 (2000).

\bibitem{RGE14} S. Antusch {\it et al.}, Phys. Lett. B {\bf 519}, 238 (2001).

\bibitem{RGE15} S. Antusch {\it et al.}, Phys. Lett. B {\bf 525}, 130 (2002).

\bibitem{RGE16} S. Antusch, J. Kersten, M. Lindner, and M. Ratz,
Nucl. Phys. B {\bf 674}, 401 (2003).

\bibitem{RGE17} S. Antusch {\it et al.}, JHEP {\bf 0503}, 024 (2005).

\bibitem{RGE18} M. Lindner, M. Ratz, and M. A. Schmidt,
JHEP {\bf 0509}, 081 (2005).

\bibitem{RGE19} Z.Z. Xing, Phys. Lett. B {\bf 633}, 550 (2006).

\bibitem{RGE20} Z.Z. Xing and H. Zhang, Commun. Theor. Phys.  {\bf 48}, 525 (2007).

\bibitem{Weinberg1979} S. Weinberg, Phys. Rev. Lett. {\bf 43}, 1566 (1979).

\bibitem{LMX05} S. Luo, J.W. Mei, and Z.Z. Xing,
Phys. Rev. D {\bf 72}, 053014 (2005).

\bibitem{Xing09} For a brief review, see: Z.Z. Xing,
Prog. Theor. Phys. Suppl. {\bf 180}, 112 (2009); and references therein.

\bibitem{LR1} V. Tello, M. Nemevsek, F. Nesti, G. Senjanovic and F. Vissani, Phys. Rev. Lett.  {\bf 106}, 151801 (2011).

\bibitem{LR2} M. Nemevsek, F. Nesti, G. Senjanovic and V. Tello, arXiv:1112.3061 [hep-ph].

\bibitem{LR3} M. Nemevsek, G. Senjanovic and V. Tello, arXiv:1211.2837 [hep-ph].

\bibitem{XZ12}  Z.Z. Xing and Y.L. Zhou, Phys. Lett. B {\bf 715}, 178 (2012).

\bibitem{CIB1} See, e.g., A. Mirizzi, D. Montanino, and P. Serpico,
Phys. Rev. D {\bf 76}, 053007 (2007).

\bibitem{CIB2} S. Matsuura {\it et al.}, Astrophys. J. {\bf 737}, 2 (2011).

\bibitem{CIB3} S.H. Kim {\it et al.}, J. Phys. Soc. Jap. {\bf 81}, 024101 (2012).

\bibitem{Xing2008} Z.Z. Xing, Phys. Lett. B {\bf 660}, 515 (2008).

\bibitem{Shrock} R.E. Shrock, Nucl. Phys. B {\bf 206}, 359 (1982).

\bibitem{Wolfenstein} P.B. Pal and L. Wolfenstein,
Phys. Rev. D {\bf 25}, 766 (1982).

\bibitem{Kayser} B. Kayser, Phys. Rev. D {\bf 26}, 1662 (1982).

\bibitem{Nieves} J.F. Nieves, Phys. Rev. D {\bf 26}, 3152 (1982).

\bibitem{Antusch} S. Antusch {\it et al.}, JHEP {\bf 0610}, 084 (2006).

\bibitem{GIM} S.L. Glashow, J. Iliopoulos, and L. Maiani,
Phys. Rev. D {\bf 2}, 1285 (1970).

\bibitem{Raffelt99} G.G. Raffelt, Phys. Rept. {\bf 320}, 319 (1999).

\bibitem{Giunti09} C. Giunti and A. Studenikin, Phys. Atom. Nucl. {\bf 72},
2089 (2009).

\bibitem{Pilaftsis} A. Pilaftsis, Z. Phys. C {\bf 55}, 275 (1992).

\bibitem{Zhou} S. Zhou, {\it PhD Thesis} (Institute of High Energy Physics,
Beijing, 2009).

\bibitem{DeGouvea} A. de Gouvea, J. Jenkins, and N. Vasudevan,
Phys. Rev. D {\bf 75}, 013003 (2007).
\end{thebibliography}
\end{document}